%% file: Main.tex
\documentclass[acmlarge,authorversion]{acmart}

\AtBeginDocument{%
  }

\setcopyright{acmcopyright}
\copyrightyear{2023}
\acmYear{2023}
\acmDOI{XXXXXXX.XXXXXXX}

\acmPrice{15.00}
\acmISBN{978-1-4503-XXXX-X/18/06}

\acmJournal{TOSEM}
\acmVolume{1}
\acmNumber{1}
\acmArticle{1}
\acmMonth{1}

\newcommand*{\eg}{e.g., }
\newcommand*{\ie}{i.e., }

\newcommand*{\toolname}{COMET\xspace}
\newcommand*{\memoo}{COMET\textsubscript{o}\xspace}
\newcommand*{\memor}{COMET\textsubscript{r}\xspace}
\newcommand*{\detectedbugs}{32\xspace}
\newcommand*{\confirmedbydevelops}{21\xspace}

\newcommand*{\fixedbugs}{7\xspace}
\newcommand*{\remainedbugs}{11\xspace}

\newcommand*{\mycode}{\fontfamily{lmtt}\selectfont}
\usepackage{caption}
\usepackage{subcaption}
\usepackage{enumitem}
\usepackage{multirow}
\usepackage{algorithmic}
\usepackage[linesnumbered,ruled,vlined]{algorithm2e}
\usepackage{color,colortbl}
\definecolor{Gray}{gray}{0.9}
\definecolor{darkgreen}{rgb}{0,0.5,0}
\usepackage{pgfplots}
\usepackage{booktabs,caption}

\usepackage{pifont}
\newcommand{\cmark}{\ding{51}}%
\usepackage[flushleft]{threeparttable}

\pgfplotsset{compat=1.17} 
\begin{document}
\title{\toolname: Coverage-guided Model Generation For Deep Learning Library Testing}
\author{Meiziniu Li}
\orcid{0000-0001-5947-4030}
\affiliation{%
  \institution{The Hong Kong University of Science and Technology}
  \country{China}
}
\email{mlick@cse.ust.hk}

\author{Jialun Cao}
\orcid{0000-0003-4892-6294}
\affiliation{
  \institution{The Hong Kong University of Science and Technology}
  \country{China}
}
\affiliation{
   \institution{Guangzhou HKUST Fok Ying Tung Research Institute}
  \country{China}
}
\email{jcaoap@cse.ust.hk}

\author{Yongqiang Tian}
\orcid{0000-0003-1644-2965}
\affiliation{
    \institution{University of Waterloo}
    \country{Canada}
}
\affiliation{
    \institution{The Hong Kong University of Science and Technology}
    \country{China}
}
\email{yongqiang.tian@uwaterloo.ca}

\author{Tsz On Li}
\orcid{0000-0001-8031-4947}
\affiliation{%
  \institution{The Hong Kong University of Science and Technology}
  \country{China}
}
\email{toli@connect.ust.hk}

\author{Ming Wen*}
\orcid{0000-0001-5588-9618}
\affiliation{%
  \institution{Huazhong University of Science and Technology}
  \country{China}
}
\email{mwenaa@hust.edu.cn}

\author{Shing-Chi Cheung*}
\orcid{0000-0002-3508-7172}
\thanks{* Corresponding author.}
\affiliation{
  \institution{The Hong Kong University of Science and Technology}
  \country{China}
}
\affiliation{
   \institution{Guangzhou HKUST Fok Ying Tung Research Institute}
  \country{China}
}
\email{scc@cse.ust.hk}

\renewcommand{\shortauthors}{Li, et al.}

\begin{abstract}

Recent deep learning (DL) applications are mostly built on top of DL libraries. The quality assurance of these libraries is critical to the dependable deployment of DL applications. Techniques have been proposed to generate various DL models and apply them to test these libraries.
However, their test effectiveness is constrained by the diversity of layer API calls in their generated DL models.
Our study reveals that these techniques can cover at most 34.1\% layer inputs, 25.9\% layer parameter values, and 15.6\% layer sequences.
As a result, we find that many bugs arising from specific layer API calls (i.e., specific layer inputs, parameter values, or layer sequences) can be missed by existing techniques.

Because of this limitation, we propose \toolname to effectively generate DL models with diverse layer API calls for DL library testing. 
\toolname: (1) designs a set of mutation operators and a coverage-based search algorithm to diversify layer inputs, layer parameter values, and layer sequences in DL models.
(2) proposes a model synthesis method to boost the test efficiency without compromising the layer API call diversity.
Our evaluation result shows that \toolname outperforms baselines by covering twice as many layer inputs (69.7\% vs. 34.1\%), layer parameter values (50.2\% vs. 25.9\%), and layer sequences (39.0\% vs. 15.6\%) as those by the state-of-the-art.
Moreover, \toolname covers 3.4\% more library branches than those by existing techniques. Finally, \toolname detects 32 new bugs in the latest version of eight popular DL libraries, including TensorFlow and MXNet, with \confirmedbydevelops of them confirmed by DL library developers and \fixedbugs of those confirmed bugs have been fixed by developers.
\end{abstract}




\maketitle

\input{1-Introduction-3}
\input{2-Background-2}

\input{5-Methodology-2}
\input{6-Evaluation}
\input{7-Discussions}

\input{8-Threats}
\input{9-RelatedWork}
\input{10-Conclusion}
\input{11-Acknowledgements}

\bibliographystyle{ACM-Reference-Format}
\bibliography{reference}

\end{document}

%% file: 1-Introduction-3.tex
\section{Introduction}
\label{sec:introduction}
Deep learning (DL) has been increasingly adopted for mission-critical applications 
such as authentication~\cite{das2018deep, Ferdowsi2019DeepLF}, medical treatment~\cite{litjens2017survey}, pandemic control~\cite{Ardakani2020ApplicationOD,Bhattacharya2020DeepLA}, and autonomous driving~\cite{autonomousdriving,Sallab2017DeepRL,ShalevShwartz2016SafeMR}. Currently, many recent works focus on assuring the quality of DL applications, e.g., testing the trained DL models~\cite{pei2017deepxplore,tensorfuzz,deeproad} or testing the DL program written by DL application developers~\cite{debar,wardat2021deeplocalize}. However, there are only a few research works on testing DL libraries, such as TensorFlow~\cite{tensorflow}, PyTorch~\cite{pytorch}, MXNet~\cite{mxnet}, and ONNX~\cite{onnx}. These DL libraries are the basis during the development and deployment of current DL applications. Nevertheless, recent studies~\cite{JIA2021110935, chen2022toward} found that these libraries suffer from coding bugs frequently. Early detection of these bugs is crucial.

Prior works~\cite{cradle,lemon,audee,muffin,graphfuzz} mainly follow the same testing paradigm, i.e., feeding various DL models (called \textit{test inputs}) to exercise specific modules (e.g., model loading, construction, and inference modules) in DL libraries.
To obtain sufficient test inputs (i.e., DL models), existing works either generate DL models by mutating the network structures~\cite{lemon,audee,graphfuzz}, parameters~\cite{lemon,audee}, or input value~\cite{cradle,lemon,audee,graphfuzz} of published models (e.g., ResNet~\cite{resnet} and InceptionNet~\cite{inceptionnet}) or build DL models from scratch~\cite{muffin} based on predefined model structure templates (e.g., chain structure~\cite{muffin}).

\begin{figure}[t!]
    \centering
    \resizebox{0.6\linewidth}{!}{
    \includegraphics{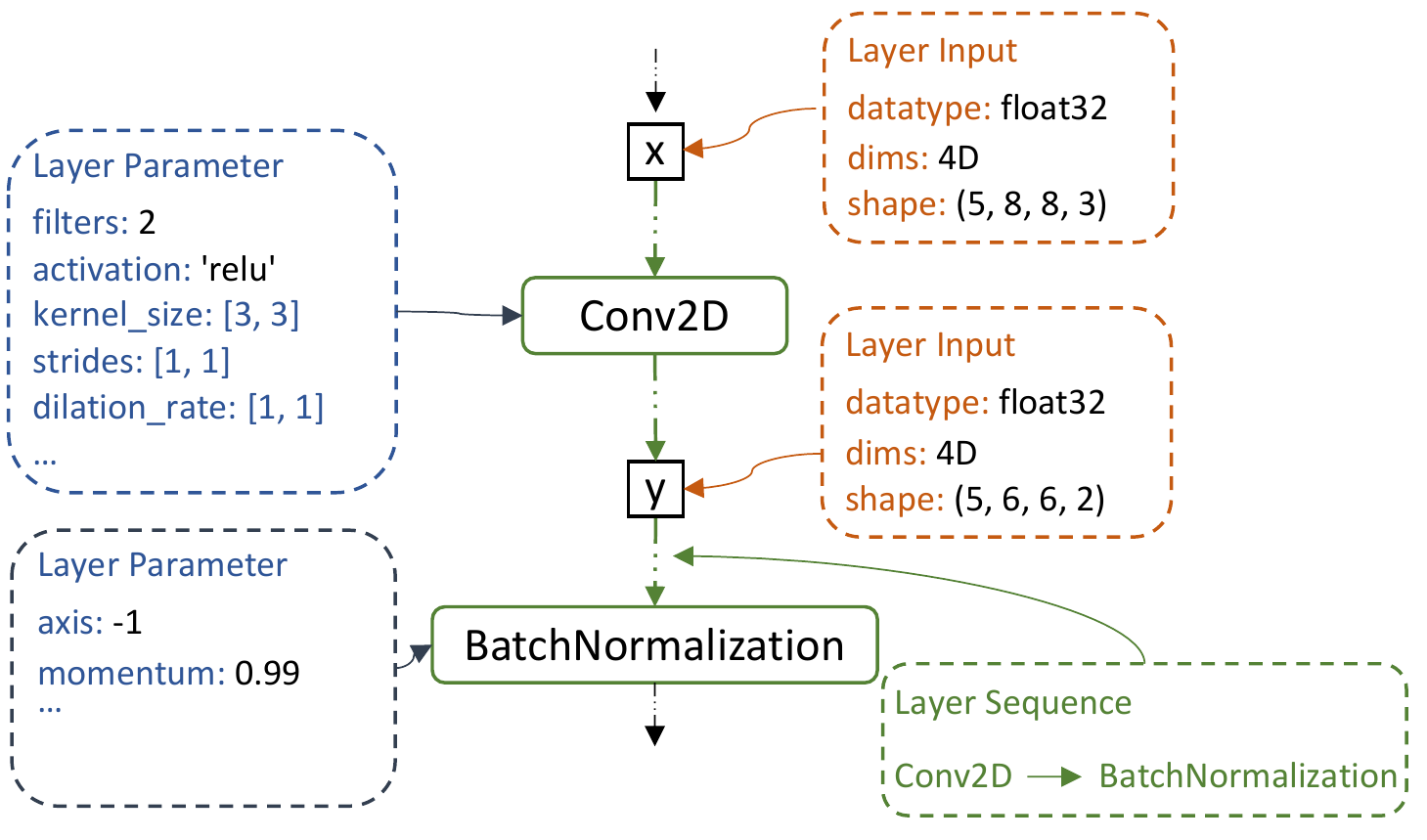}
    }
\caption{An Example of A Layer API Call In DL Model}
    \label{fig:model_diversity_example}
\end{figure}

However, the test adequacy of existing works is far from satisfactory (\eg~achieving low test coverage~\cite{chen2022toward}).
According to our investigation (see \S\ref{sec:bg_limitation}), only 16\% of branches in TensorFlow's model construction and model execution modules could be covered even using the latest testing technique.
The unsatisfactory result reveals the inadequacy of existing techniques. Taking a closer look, we observed that a main reason is the lack of diversity of test inputs (i.e., DL models). 
Specifically, existing techniques target generating DL models to either enlarge the prediction inconsistency in differential testing~\cite{lemon,audee}, increase the coverage of layer APIs~\cite{muffin} or increase the graph structure diversity~\cite{graphfuzz}. However,
they are weak in generating DL models to expose bugs that reside in some specific layer API calls. 
Take Figure~\ref{fig:model_diversity_example} as an example. The diversity of layer API calls inside a DL model is attributed to three finer properties: (1) \textit{layer inputs} (e.g., input datatype and number of input dimensions, abbreviated to \textit{dims}, highlighted in orange), (2) \textit{layer parameter values} (e.g., values of {\mycode filters} and {\mycode kernel\_size} highlighted in blue), and (3) \textit{layer sequences} (e.g., the connection between {\mycode Conv2D} layer and {\mycode BatchNormalization} layer highlighted in green). 
However, neither covering more types of layer APIs~\cite{muffin} nor diversifying the graph structure~\cite{graphfuzz} could encourage generating layer API calls with diverse layer parameter values or layer inputs. Indeed, we found that models generated by existing techniques~\cite{cradle,lemon,muffin} do not cover most layer API calls, i.e., at most 34.1\% of layer inputs, 25.9\% of layer parameter values, and 15.6\% of layer sequences could be covered (see \S\ref{sec:bg_limitation}).

\begin{figure}[t!]
    \centering
    \resizebox{0.4\linewidth}{!}{
    \includegraphics{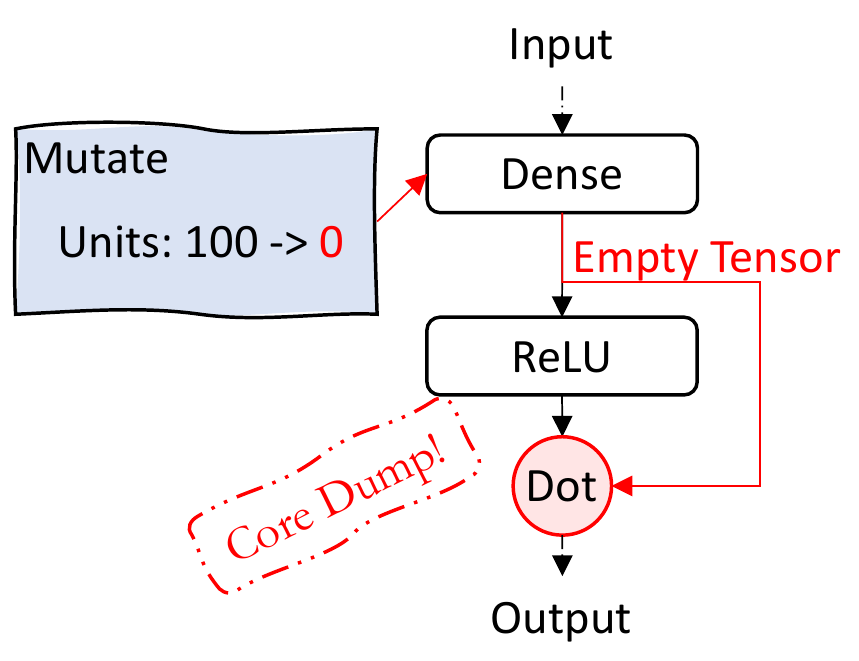}
    }
\caption{A Crash Bug of ONNXRuntime}
    \label{fig:intro_example}
\end{figure}

Consequently, existing techniques cannot effectively detect bugs residing at specific layer API calls. Figure~\ref{fig:intro_example} shows a real bug detected by us that illustrates the weaknesses of these techniques.\footnote{\url{https://github.com/microsoft/onnxruntime/issues/11024}}
It is a crash bug triggered by a variant of the ResNet model, leading to a core dump in ONNXRuntime, a famous DL library for model inference and training with over 6.5K stars in GitHub. 
The bug can only be triggered when models with a specific layer sequence (i.e., {\mycode Dense}$\rightarrow${\mycode Dot}) and layer input (as shown on the left with the layer input of {\mycode Dot} layer to has zero shapes) are generated. 
Moreover, we found out that 18 out of \confirmedbydevelops confirmed new bugs detected by our tools are manifested by some specific layer API calls.
In other words, testing techniques designed for diversifying layer API calls are desired for testing DL libraries.

To this end, we propose \toolname, a \textbf{CO}verage-guided \textbf{M}odel gen\textbf{E}ration tool for DL library \textbf{T}esting. \toolname is designed to generate DL models with diverse layer API calls.
Similar to LEMON~\cite{lemon} and Audee~\cite{audee}, \toolname constructs the initial seeds based on published DL models. It then generates new DL models by iteratively mutating them.
The challenges of designing \toolname are two-fold. First, \textbf{it is hard to generate DL models with diverse layer API calls} due to the huge search space.
For instance, considering a widely-used network, InceptionResnetV2, with 782 layers and 30,400 layer parameters, by simply considering each possible layer parameter once, it will take more than 30 thousand times to test, let alone considering the model structure. Therefore, random mutation (e.g., randomly choosing a layer into a random position inside a DL model), as is used by existing techniques~\cite{lemon,audee,graphfuzz}), may derive a layer API call that previously generated models has already covered, thus is of no benefit for diversifying layer API calls. All in all, it is impossible to exhaustively search for all possible cases in the huge search space with respect to a model. 
Second, \textbf{the runtime overhead when generating and executing a DL model is expensive}. 
As reported by an earlier work~\cite{freelunch}, testing techniques based on model generation and execution can be cost-prohibitive.
It can take over 10 minutes to process a new DL model based on InceptionResNetV2 on an Intel GPU server, which poses a significant challenge to test efficiency. Iteratively mutating these published models, mostly large, incurs non-trivial runtime overhead in model generation.

The insight of \toolname to address these two challenges is based on the observation: generating DL models with new layer inputs, parameter values, or sequences is likely to invoke different layer API calls, thus achieving more comprehensive testing on DL libraries.
Driven by the insight, we propose \toolname, which mainly contains three novelties.  
First, we design a set of mutation operators\footnote{Different from mutation testing works~\cite{deepmutation,tensorfuzz,deepcrime,adversamplemutation} that insert error into DL models and consider them as subject under test. Our work uses mutation operators to generate new DL models and we further uses the generated DL models as test inputs for DL library testing.} aiming to target diversifying layer API calls by mutating the layer inputs, layer parameter values, and the layer sequence inside the DL model. Instead of random mutation, our mutation operators search for the optimal mutation option to target mutating DL models with diverse layer API calls. Second, since not all mutation operators are equally effective in diversifying DL models and increasing test coverage, we employ the Monte Carlo Markov Chain (MCMC) algorithm with the coverage-based fitness function to iteratively mutate seed models to achieve higher layer API call diversity and test coverage.
Third, we design a model synthesis method for published models to reduce their size while preserving the diversity of layer inputs, parameters, and sequences. Benefiting from the model synthesis method, \toolname can reduce the runtime overhead during model generation and execution. 

We evaluate the performance of \toolname on eight popular DL libraries: Keras~\cite{keras}, TensorFlow~\cite{tensorflow}, PyTorch~\cite{pytorch}, MXNet~\cite{mxnet}, ONNXRuntime~\cite{onnxruntime}, Keras-MXNet~\cite{keras-mxnet}, TF2ONNX~\cite{tf2onnx}, ONNX2PyTorch~\cite{onnx2pytorch}. 
In particular, we adopt differential testing for bug detection, i.e., the inconsistencies are compared between DL libraries. While the DL model's implementation varies from library to library, a model conversion step is thus needed to enable comparison. 
Therefore, the first five libraries (\ie~Keras, TensorFlow, PyTorch, MXNet, and ONNXRuntime) are tested using differential testing, exercising their model construction and inference modules. While the rest libraries (\ie~Keras-MXNet, TF2ONNX, ONNX2PyTorch) are used for model conversion, thus their conversion modules are tested.
We compare \toolname with three state-of-the-art DL library testing techniques as baselines: CRADLE~\cite{cradle}, LEMON~\cite{lemon},  
and the latest technique: Muffin~\cite{muffin}. For GraphFuzz~\cite{graphfuzz} and Audee~\cite{audee}, since no executables or source code are available, we do not include them as baselines. 
The evaluation results show that \toolname's mutation operators and search algorithm can significantly increase the diversity of layer API calls in the generated models, outperforming baselines by covering 35.6\% more layer inputs, 24.3\% more parameter values, and 23.4\% more sequences.
Moreover, \toolname can cover at least 3.4\% more branches than baselines in TensorFlow's model construction and model execution modules.
\toolname successfully finds \detectedbugs new DL library bugs. So far, \confirmedbydevelops of them have been confirmed by DL library developers, and \fixedbugs of the confirmed ones have already been fixed by developers in the latest version, the remaining \remainedbugs unconfirmed bugs are pending review.

In summary, our work makes the following major contributions{:}
\begin{itemize}
    \item \textbf{Originality.} We propose \toolname, a diversity-driven model generation framework for DL library testing, which can improve the test coverage of existing techniques by increasing the diversity of layer API calls inside DL models.
    Specifically, we propose three coverage criteria for layer API call diversity. We further propose a set of mutation operators and a coverage-based MCMC search algorithm to diversify layer API calls. In addition, we design a model synthesis method based on our coverage criteria to boost the efficiency of \toolname significantly.
    
    \item \textbf{Effectiveness.} Our evaluation shows that \toolname outperforms the state-of-the-art by covering 35.6\% more layer inputs, 24.3\% more layer parameters, 23.4\% more layer sequences, and 3.4\% more library branches.

    \item \textbf{Usefulness.} We applied \toolname to eight popular DL libraries. It successfully detected \detectedbugs new bugs that caused crashes, NaN, and inconsistent outputs. \confirmedbydevelops out of these \detectedbugs new bugs have been confirmed by the developers. \fixedbugs of those confirmed bugs have been fixed. Our implementation of \toolname, the details of detected bugs together with experiment results are publicly available at our project site: \url{https://github.com/maybeLee/COMET}.
\end{itemize}

%% file: 2-Background-2.tex
\section{Background and Motivation}
\label{sec:bg}

\subsection{Testing DL Libraries by DL Models}
\label{sec:library_testing}
\begin{figure}[t!]
    \centering
    \resizebox{0.8\linewidth}{!}{
    \includegraphics{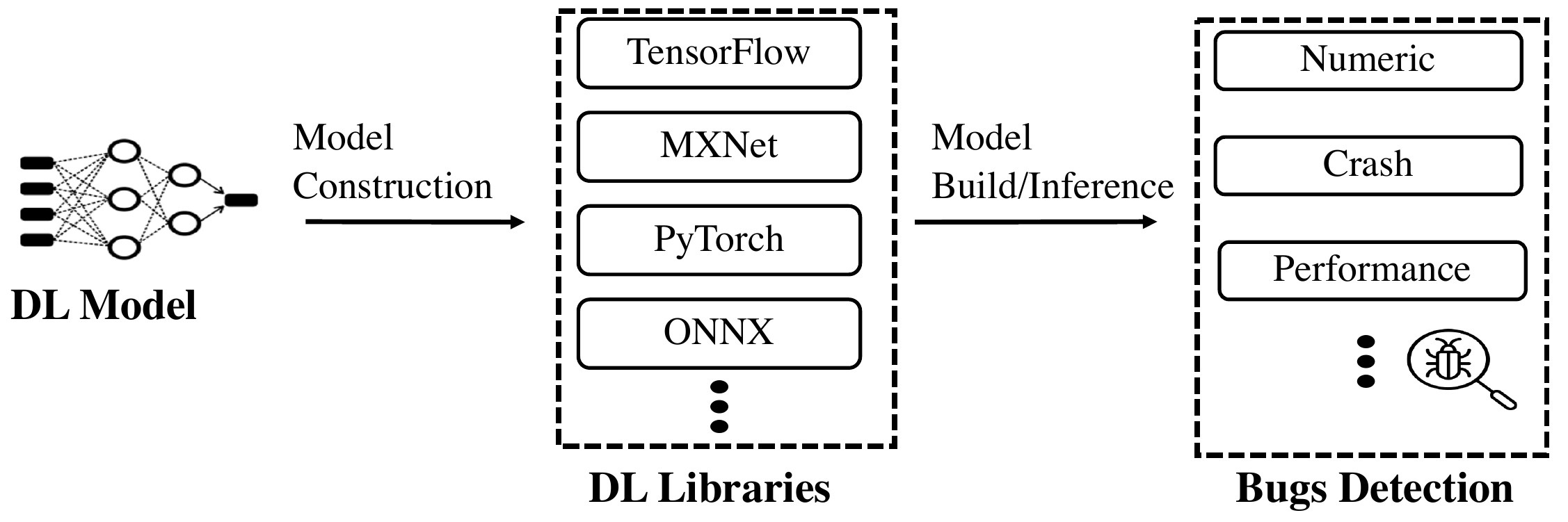}
    }
\caption{Overview of Testing DL Library by DL Models.}
    \label{fig:test_patterns}
\end{figure}

Prior testing techniques commonly use DL models as test input to test DL libraries.
As illustrated in Figure~\ref{fig:test_patterns}, feeding these models to DL libraries for making inferences will execute code in the loading and inferencing modules. For test oracle design, they adopt differential testing to apply the same DL model on multiple DL libraries with model conversion tools or shared high-level APIs. Inconsistent behavior (i.e., inconsistent outputs or inconsistent symptoms between DL libraries) indicates potential library bugs. They further design a set of mutation operators or generation rules to obtain sufficient test inputs (i.e., DL models). These techniques complement the unit test suites of a DL library by invoking a sequence of library APIs for the end-to-end completion of machine learning.

CRADLE~\cite{cradle} is the earliest technique proposed to test DL libraries by collecting publicly-available models and datasets to detect inconsistent outputs across DL libraries. It successfully detects implementation differences of Keras across different library backends.
Based on these existing DL models, two model generation techniques, LEMON~\cite{lemon} and Audee~\cite{audee}, were later proposed to generate DL models driven by output inconsistency or NaN symptoms. These two techniques present a set of mutation operators, such as mutating the model's weights, architectures, and layer parameters to explore library code. For guiding model mutation, they either use the inconsistency of generated models' outputs over different DL libraries as the guidance to expose the incorrect computation or use a heuristic-based fitness function to encourage mutating DL models that tend to output outliers (i.e., too large or small values) to expose NaN values. GraphFuzz~\cite{graphfuzz} is further proposed to mutate the computation graph in the DL model for DL library testing. It introduces an operator-level coverage to quantify the model diversity and further designs a set of mutation operators to explore combinations of model structures and parameters. Muffin~\cite{muffin} is the most recent DL library testing work that generates DL models driven by layer diversity. In particular, Muffin abstracts the DL model as a Directed Acyclic Graph (DAG) and defines two types of commonly used graph structures: chain structure and cell-based structure to generate the structure information (i.e., the topology of how layers are connected); it further generates DL models towards increasing the coverage of layer APIs.

\subsection{Model Generation Strategies}
\label{sec:bg_objectives}

Existing techniques, as mentioned in \S\ref{sec:library_testing}, adopt different strategies in model generation. For example, LEMON generates models to enlarge the output inconsistency across DL libraries. Besides output inconsistency, Audee also favors generating models that can trigger the Not-A-Number (NaN) value in DL libraries.\footnote{Bugs detected by Audee are available at the site: \url{https://sites.google.com/view/audee/bug-details}.} GraphFuzz designs an operator-level coverage defined as a sum of five coverage criteria (i.e., operator type coverage, input degree coverage, output degree coverage, single edge coverage, shape\&parameter coverage) derived from graph theory. It further iteratively mutates models towards a higher operator-level coverage result.\footnote{Exceptions detected by GraphFuzz are available at the site: \url{https://github.com/gbftdlie/Graph-based-fuzz-testing}.}
Muffin adopts the strategy of maximizing layer type coverage (i.e., the percentage of invoked layer APIs in all the pre-defined layer APIs) in model generation.

When testing DL libraries using DL models, the code executed in DL libraries is related to the layer API calls (as is exemplified in Figure~\ref{fig:model_diversity_example}) used to define the DL models. \textbf{First, diversifying layer input can help better test DL libraries.}
Specifically, changing the datatype of layer input can test those precision-specific codes in DL libraries~\cite{wang2016precision}. For instance, in TensorFlow, changing the input datatype of {\mycode BatchNormalization} ( or {\mycode LayerNormalization}) from {\mycode float32} to {\mycode float16} (or non-{\mycode float32}) will trigger new branches handling numeric instability (or layer fusion) issues. Changing the number of dimensions for a tensor will change the data formats it represents (e.g., 3D tensors usually refer to text data, and 4D tensors represent image data), thus may test new code in DL libraries. For example, changing the input of {\mycode BatchNormalization} layer from a {\mycode 4D} tensor to a {\mycode 5D} tensor will trigger new functions related to operator fusion during the model construction.
Changing the shape of layer input may also affect library code, such as codes related to weights initialization and tensor padding during the model construction. An example of this observation is that different tensor input shapes may affect how the tensor is padded~\cite{tensorflow_padding}. Besides, our experiment also found the manifestation of some bugs required a specific layer input shape during the model conversion~\cite{onnx2pytorch_41}.
\textbf{Second, diversifying layer parameter values can also help increase the test coverage.} For instance, changing the value of {\mycode activation} from {\mycode relu} to {\mycode sigmoid} will test a different activation function in {\mycode Conv2D}. 
Changing the value of {\mycode strides} and {\mycode dilation\_rate} may result in the incorrect padding logic inside the {\mycode Conv2D} layer, which is a confirmed bug~\cite{keras_issue_16314} we detected. 
Setting {\mycode units} of the {\mycode Dense} layer to 0 will generate an empty-shape tensor which may result in a direct core dump if this empty-shape tensor is directed to a {\mycode Conv3DTranspose} layer~\cite{keras_issue_16933}. 
Changing the value of {\mycode kernel\_size} will also influence how {\mycode Conv2D} does the padding when constructing the model~\cite{keras_kernel_size_padding}.
\textbf{Third, testing DL libraries with different layer sequences can also help to explore library code.}
On the one hand, introducing a layer sequence with new layer types can test more layer APIs~\cite{muffin}. On the other hand, covering some specific layer sequences may also cover new codes related to model construction and model computation inside DL libraries. For instance, the layer sequence {\mycode Conv2D}$\rightarrow${\mycode ReLU} will trigger the operator fusion for Conv2D and Mul operator~\cite{onnxruntime_conv_fusion} and the {\mycode FindContractionWithBias} function in TensorFlow. 

However, existing model generation strategies are weak in diversifying layer API calls in DL models. First, enlarging output inconsistency has little help in diversifying layer API calls. Second, only focusing on layer type coverage can offer no guidance for diversifying other properties, i.e., layer parameter values, layer inputs, and layer sequences.
Although the operator-level coverage proposed by GraphFuzz provides a more optimal granularity in guiding model generation to increase layer sequence diversity, they are still weak in diversifying layer parameters and layer inputs. Therefore, some library code related to some specific layer API calls cannot be covered by existing techniques.
We observed that missing specific layer API calls may fail to detect some library bugs. 
This observation motivates us to define more optimal criteria for measuring the layer API call diversities, including \textbf{layer sequence diversity} adapted from existing works~\cite{muffin,graphfuzz} and two newly proposed ones: \textbf{layer input diversity} and \textbf{layer parameter diversity}. Specifically, layer input diversity includes the variety of input's datatype (e.g., {\mycode float16}), input dimension (e.g., {\mycode 4D}), and input shape (e.g., {\mycode (batch\_size,24,24,3)});
layer parameter diversity includes the diversities of parameter values for each type of layer;
layer sequence diversity comprises the flow of tensor from one layer API to another.
All these three diversities collectively contribute to the diversity of layer API calls. Accordingly, the core objective of our paper is to better test DL libraries by increasing the diversity of layer inputs, parameters, and sequences in the generated models; thus, we can achieve higher test coverage.

\subsection{Limitations of Existing Techniques}
\label{sec:bg_limitation}
To explore the limitations of baselines~\cite{cradle,lemon,muffin}, we experimented with measuring their performance in diversifying layer API calls. In particular, we use our proposed three coverage criteria: layer input coverage, layer parameter coverage, and layer sequence coverage (see \S\ref{subsec:test-space}) as the metrics.
Additionally, we record the branch coverage on TensorFlow's model construction and model execution modules
achieved by existing works as the representative to reveal their inadequate testing effectiveness. We run each technique with the default hyperparameters for six hours and then analyze the coverage results.
The experiment results show low coverage with respect to all four coverage criteria, suggesting the limited effectiveness of existing works in diversifying layer API calls and achieving satisfactory test coverage on DL libraries. Specifically, only at most 34.1\% of layer inputs (261/766), 15.6\% of layer sequences (338/2170), and 25.9\% of layer parameter values (530/2049) have been covered by existing techniques. 
As a result, only at most 16.0\% of library branches are covered. 
Moreover, we found that bugs (like Figure~\ref{fig:intro_example}) residing at some specific layer API calls and library branches are not detected by existing techniques. However, these bugs may lead to severe consequences (e.g., a core dump), which may affect the reliability of DL libraries.
Motivated by this finding, we propose \toolname to generate diverse models to increase the test coverage for DL library testing.

\begin{table}[htbp!]
\caption{Various Coverages by Existing Techniques}
\label{tab:mot_test_coverage}
\renewcommand{\arraystretch}{1.0}
\setlength{\tabcolsep}{2.2pt}
\resizebox{0.8\linewidth}{!}{
\begin{tabular}{l|l|l|l|l}
\hline
\multirow{2}{*}{Technique} & Layer Input & Layer Parameter & Layer Sequence & Branch\\
 & Coverage & Coverage & Coverage & Coverage\\

\hline\hline
MUFFIN
& 34.1\%
& 25.9\%
& 15.6\%
& 16.0\% \\ \hline 
\rowcolor{Gray}
LEMON
& 14.2\%
& 6.4\%
& 2.4\%
& 12.5\%  \\ \hline
CRADLE
& 10.4\%
& 5.9\%
& 1.2\%
& 12.0\%\\ \hline
\end{tabular}
}
\end{table}

%% file: 5-Methodology-2.tex
\section{Methodology}
\label{sec:methodology}
\subsection{Overview}\label{subsec:overview}

\begin{figure*}[htbp!]
    \centering
    \includegraphics[width=1\linewidth]{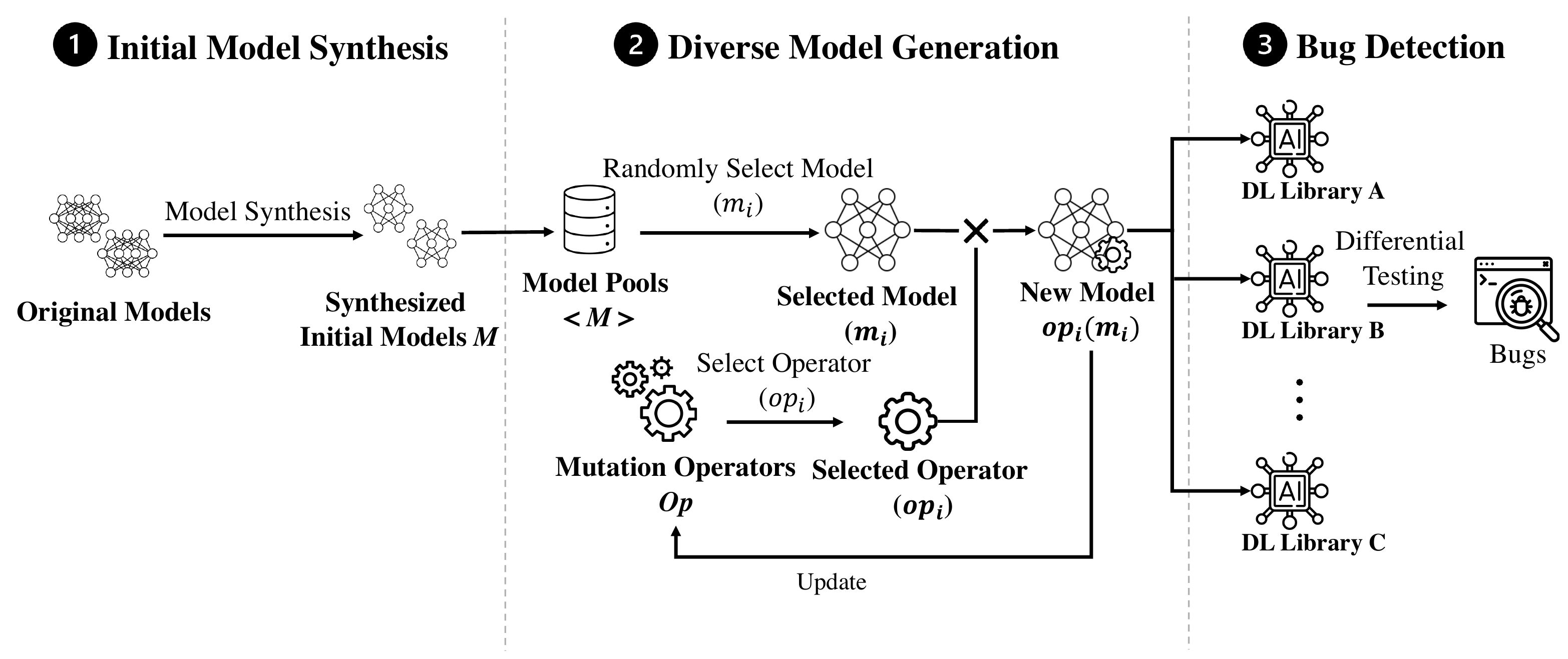}
    \caption{Overview of \toolname}
    \label{fig:methodology_overview}
\end{figure*}

Our work, \toolname, is designed to generate DL models with diverse layer API calls effectively. As defined earlier, a layer API call consists of three properties: layer input, parameter values, and sequences. We first propose three coverage criteria (in \S\ref{subsec:test-space}) for the layer API call to measure its diversity. Based on our coverage criteria, we design the workflow of \toolname into three steps (Figure~\ref{fig:methodology_overview}): (1) \textit{Initial Model Synthesis}, (2) \textit{Diverse Model Generation}, and (3) \textit{Library Testing}. We first apply our model synthesis (\S\ref{sec:methodology-reduction}) method on published DL models to synthesize the initial seeds with small size while retaining the published DL models' layer API call diversity. We further apply our mutation operators (\S\ref{sec:methodology-mutation}) and MCMC algorithm (\S\ref{sec:methodology-mcmc}) to iteratively generate models toward higher layer API call diversity and higher test coverage in DL libraries. Finally, we use these generated models as test input for library testing (\S\ref{subsec:bug_detection}).

\subsection{Coverage Criteria For layer API calls}
\label{subsec:test-space}
Motivated by Figure~\ref{fig:model_diversity_example}, we focus on three properties (i.e., layer input, layer parameter values, and layer sequences) to measure the diversity of layer API calls. The space of these properties is summarized in Table~\ref{tab:model_diversity_space}. Specifically, a diverse layer API call is found if a DL model contains new \textit{layer inputs}, \textit{parameter values}, or \textit{sequences} that previously generated models have not covered. Based on these three properties, we further introduce three coverage criteria (i.e., layer input coverage, layer parameter coverage, and layer sequence coverage) to measure the parts of layer API calls exercised by the generated models.

\input{Table-Diversity-Space}

\subsubsection{\textbf{Layer Input Coverage}}
A layer input is a tensor (i.e., a multi-dimensional array).
The datatype, number of dimensions, and shape (i.e., the number of elements in each dimension)
are three essential properties that determine an array. As exemplified by the tensor {\mycode x} in Figure~\ref{fig:model_diversity_example}, the datatype determines the floating point format stored in the computer (e.g., {\mycode float32} means {\mycode x} is stored in the single-precision format). The number of dimensions (abbreviated to ``dims'') of {\mycode x} is 4D means that {\mycode x} contains four dimensions. Usually, dimensions represent different meanings in DL libraries (e.g., in TensorFlow, the four dimensions in {\mycode x} refer to the batch size, height, width, and the number of channels, respectively). The shape of {\mycode x} determines the number of elements in each dimension (e.g., {\mycode x} is a three-channel tensor with the length of width and height to be 8). 
Motivated in \S~\ref{sec:bg_objectives}, mutating the \textit{datatype}, \textit{number of dimensions}, or \textit{shape} for a layer's input may trigger different usage of library code.
Therefore, for an input of a layer, we capture these three properties: \textit{input datatype}, \textit{input dimension}, and \textit{input shape}. However, the total number of possible values for each property is already huge (e.g., the state-of-the-art~\cite{muffin} can cover only 10.7\% input datatype, 41\% input dimension, and 34.1\% input shape). Exhaustive testing that also considers the interaction between these properties will significantly boost the test space~\cite{kuhn2004interaction}, and it may be expensive to find bugs
related to only a specific value of input datatype, input dimension, or input shape. 
Therefore, our layer input coverage encourages covering all possible values for each property rather than testing all combinations.
Accordingly, we define our layer input coverage for each layer type $l$ using Equation~\ref{eq:input_coverage}. Given the number of covered datatypes ($cov_{type}$), dimensions ($cov_{dim}$), shapes ($cov_{shape}$), and the total number of possible datatypes ($n_{type}$), dimensions ($n_{dim}$), and shapes ($n_{shape}$), the layer input coverage $Cov_{I}$ measures the portion of values covered for each property. 

\begin{align}
    Cov_{I}(l) = \frac{cov_{type}(l)+cov_{dim}(l)+cov_{shape}(l)}{n_{type}(l)+n_{dim}(l)+n_{shape}(l)}
    \label{eq:input_coverage}
\end{align}

\subsubsection{\textbf{Layer Parameter Diversity}}
\label{subsubsec:layer_parameter_diversity}
To measure the layer parameter diversity, for each layer type $l$, we record the possible parameter values $v$ for each parameter $p$ inside $l$. Layer parameters can be briefly grouped into two categories based on the type of the parameter value: numeric parameter and non-numeric parameter. Numeric parameter refers to those that require a number or a list of numbers as parameter values, e.g., the {\mycode rate} for the {\mycode Dropout} layer, {\mycode kernel\_size} for the {\mycode Conv2D} layer). Non-numeric parameter refers to those that require non-numeric values such as string or boolean as parameter values, e.g., {\mycode padding} for the {\mycode Conv2D} layer and {\mycode activation} for the {\mycode LSTM} layer. Potential parameter values for non-numeric parameters are usually categorical for each layer and can be listed following the library documentation. 
In contrast, the numeric parameter for each layer often has a huge value space. 
Since enumerating all possible values for a numeric parameter is impossible and may not help to test DL libraries, we approximate the size of the value space using a finite integer $\sigma$. Accordingly, for layer API $l$, given the number of covered parameter values $cov_{pi}$ and the total number of possible parameter values $n_{pi}$ (for numeric parameters, $n_{pi}=\sigma$) for each parameter $p_i$ inside a layer API $l$, we define the layer parameter coverage $Cov_{P}$ for layer $l$ in Equation~\ref{eq:parameter_coverage}, where $N$ is the total number of mutable parameters in $l$. In our experiment, we set $\sigma=5$.

\begin{align}
    Cov_{P}(l) = \frac{\sum^N_{i=1} cov_{pi}}{\sum^N_{i=1} n_{pi}}
    \label{eq:parameter_coverage}
\end{align}

\subsubsection{\textbf{Layer Sequence Diversity}}
Layer sequence diversity defines the execution sequence of layer APIs inside the DL model. Following GraphFuzz~\cite{graphfuzz}, we denote the layer pair as <$layer_i$, $layer_j$>, indicating the tensor is first sent to $layer_i$, and then the output of $layer_j$ is sent into $layer_j$. In other words, the space of layer API pairs is all possible permutations between all layer APIs. Note that not all layer sequences are valid. For instance, the output of the {\mycode Conv3D} layer is always a 5D tensor, and this tensor cannot be sent to the {\mycode Conv2D} layer, which only accepts a 4D tensor. We refer to this constraint as the dimension constraint (see Figure~\ref{fig:model_constraint}). In other words, any layer sequence that violates this dimension constraint is invalid. To construct the valid layer sequence, we manually collect the set of the possible number of input dimensions (denoted as $s_{idim}(l)$) and output dimensions (denoted as $s_{odim}(l)$) for each layer $l$. Finally, given the total number of layer types is $N_{layer}$ and the total number of covered layer sequences $cov_{seq}$ (note that all covered layer sequences are valid), we define the layer sequence coverage $Cov_{S}$ as below:

\begin{align}
    Cov_{S} = \frac{cov_{seq}}{\sum_{i\in [1, N_{layer}]}{\sum_{j\in [1, N_{layer}]}{\text{sign}(|s_{odim(layer_i)} \cap s_{idim(layer_j)}|)}}}
    \label{eq:sequence_coverage}
\end{align}

Specifically, we permute all the possible layer sequences and check if they are valid. For each layer sequence ($layer_i$, $layer_j$), we consider it to be valid if $layer_i$ has at least one output tensor whose number of dimensions is accepted by $layer_j$ (i.e., $s_{odim(layer_i)} \cap s_{idim(layer_j)} \neq \varnothing$). 

\subsection{Diversity-Driven Mutation}
\label{sec:methodology-mutation}
Existing model mutation operators~\cite{lemon,deepmutation,audee,graphfuzz} are mainly designed to generate mutants by changing the structure or weights of seed DL models. The design, however, does not consider the diversity of layer API calls inside DL models.
Examples of these operators are inverting the activation state of neurons and layer duplication. However, these mutation operators fail to introduce the new layer API call (i.e., new layer inputs, layer parameters, or layer sequences).
Therefore, we propose eight mutation operators in this section to generate diverse mutants for the three coverage criteria.
As shown in Table~\ref{tab:method_new_mutation}, the mutation operators could be categorized into three levels, i.e., input-level mutation, parameter-level mutation, and structure-level mutation.
In addition, we include one anomaly mutation for robustness testing.

\begin{figure}[htbp!]
    \centering
    \includegraphics[width=0.7\linewidth]{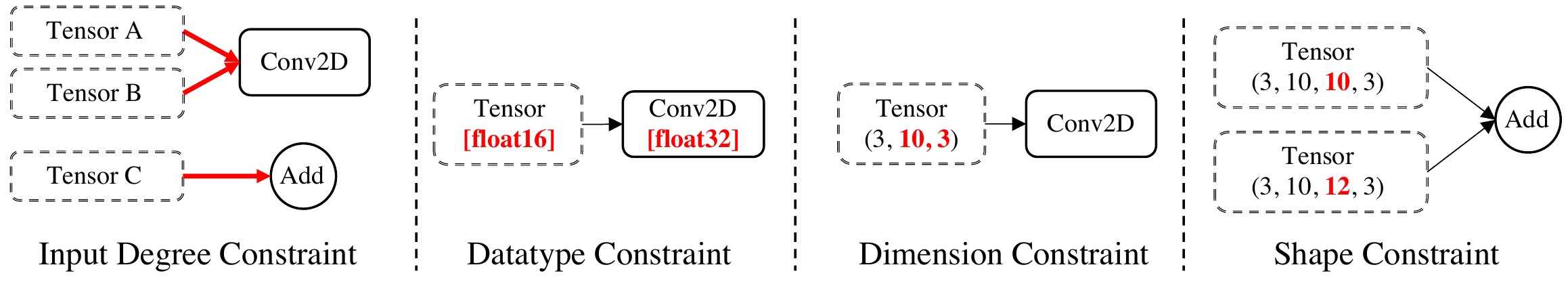}
    \caption{Example of Invalid Model Violating Tensor Constraints}
    \label{fig:model_constraint}
\end{figure}

Our mutation operators are designed to increase the diversity of layer API calls by introducing new layer inputs, parameter values, and layer sequences. Unfortunately, arbitrarily mutating DL models can easily cause model failure~\cite{muffin,lemon,graphfuzz}. In particular, model mutations must comply with some tensor constraints (i.e., rules) to avoid breaking validity checks in DL libraries. Specifically, we summarize four general tensor constraints (as exemplified in Figure~\ref{fig:model_constraint}), including three followed by existing works~\cite{lemon,muffin,graphfuzz}: input degree constraint, dimension constraint, and shape constraint, and an additional one required by us when mutating the layer input datatype: datatype constraint. 
\textbf{Input degree constraint} refers to the required number of input tensors for each layer, e.g., the {\mycode Conv2D} layer can only receive one tensor while the {\mycode Add} layer requires more than one tensor. 
\textbf{Dimension constraint} refers to the number of tensor dimensions required for each layer, e.g., the {\mycode Conv2D} layer can only accept the four-dimensional (4D) tensor, while {\mycode BatchNormalization} can accept 3D, 4D, or 5D tensor. 
We refer the \textbf{shape constraint} to the case where merging layers (e.g., {\mycode Add} layer) require a specific tensor shape (e.g., the {\mycode Add} layer requires the input tensors to have the same shape). 
Additionally, inconsistent computation precision (i.e., \textbf{datatype constraint}), determined by the datatype property, will also cause model failure. Specifically, DL libraries will directly raise an exception if the precision used for layer initialization is inconsistent with the layer input. For instance, a {\mycode Conv2D} layer initialized in 32 bits floating point ({\mycode float32}) will reject a {\mycode float16} input tensor.
The model mutation that violates either of these four constraints will result in an invalid DL model, which cannot be correctly built and loaded by DL libraries, thus, are ineffective in testing DL libraries. Therefore, we design our mutation operators to increase layer API calls' diversity while not violating four tensor constraints mentioned above.

\input{Table-New-Mutation-Operators}

\subsubsection{\textbf{Input-Level Mutation}}
\label{alg:input-mutation}

\begin{figure}[t!]
    \centering
    \resizebox{0.9\linewidth}{!}{
    \includegraphics{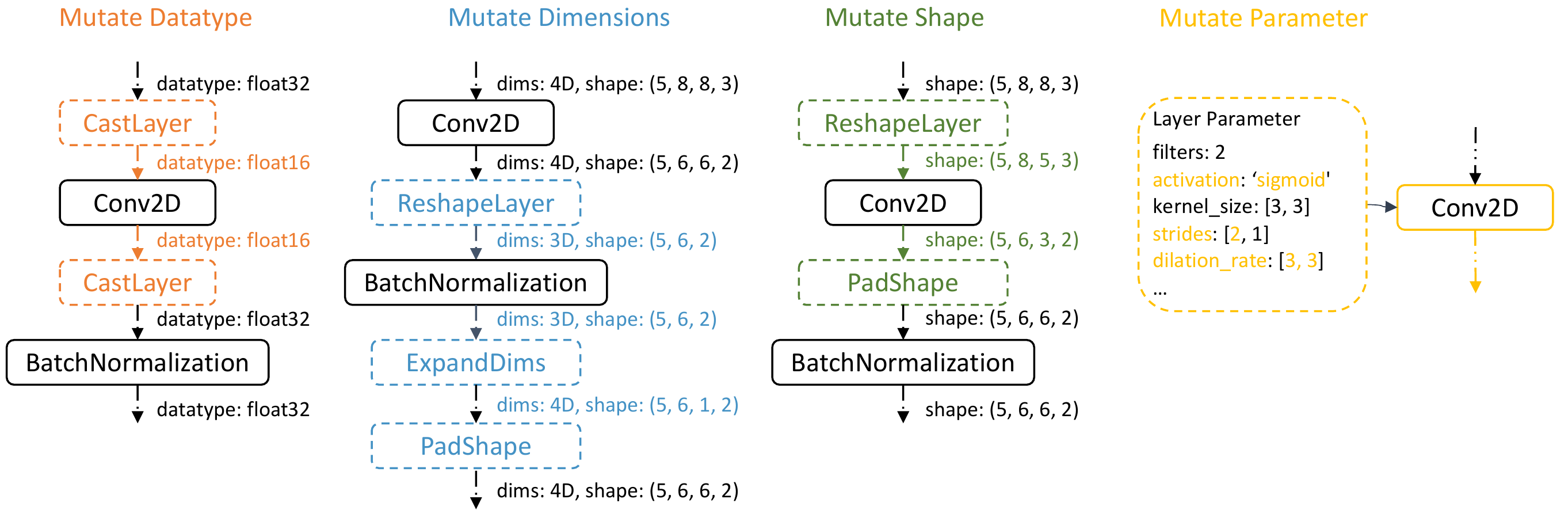}
    }
\caption{Example of Input-Level Mutation (Left Three) and Parameter-Level Mutation (Right One)}
    \label{fig:example_input_mutation}
\end{figure}

The design of input-level mutation aims to mutate the layer input's properties (i.e., datatype, number of dimensions, and shape) inside DL models. However, because of the huge search space in the layer input, randomly mutating these properties is weak in exploring new datatype, number of dimensions, and shape. Intuitively, for the target layer input to be mutated, if previously generated models have not covered a specific option (i.e., a particular input datatype, a particular number of dimensions, or a particular shape), our input-level mutation should have a higher chance to generate mutants with these options. Moreover, to increase the mutation ratio, Instead of randomly choosing one layer to mutate, our input-level mutation will randomly choose $n$ layers (denoted as $\{layer_1, layer_2, ..., layer_n\}$), where $n$ is a random integer between 1 and 10. Based on these intuitions, we design our input-level mutation, i.e., \textit{MDtype}, \textit{MDims}, and \textit{MShape}, as follows:

\begin{itemize} 
    \item \textbf{Mutate Datatype (MDtype)}: 
    Given a seed model and a list of layers $\{layer_1,..., layer_n\}$, \textit{MDtype} searches through all generated models and chooses the datatype $dtype_m$ ($m\in [1,n_{type}(layer_i)]$) that have not been covered by inputs of $layer_i$ ($i\in [1,n]$). If all datatypes have been covered, \textit{MDtype} randomly chooses one instead. Finally, for each $layer_i$, \textit{MDtype} changes its input datatype to $dtype_m$ by type casting (e.g., cast the datatype $layer_i$'s input from {\mycode float16} to {\mycode float64}). 
    \item \textbf{Mutate Dimensions (MDims)}: 
    Given a seed model and a list of layers $\{layer_1,..., layer_n\}$, \textit{MDims} searches through all generated models and chooses the possible number of dimensions $ndim_i$ that can be accepted by $layer_i$ and have not been covered by inputs of $layer_i$. If all numbers of dimensions have been covered, \textit{MDims} randomly choose one instead. Finally, for each $layer_i$, \textit{MDims} changes the number of its input dimensions to $ndim_i$ by inserting dimensions of length 1 (e.g., expand $layer_i$'s input shape from {\mycode \{10,3,3\}} to {\mycode \{10,3,3,1\}}) or randomly dropping dimensions (e.g., mutate $layer_i$'s input shape from {\mycode \{10,3,3\}} to {\mycode \{10,3\}}). 
    
    \item \textbf{Mutate Shape (MShape)}: Given a seed model and a list of layers $\{layer_1,..., layer_n\}$, \textit{MShape} searches through all generated models and checks if the number of different input shapes of $layer_i$ is larger than $n_{shape}$ (i.e., five as defined in \S\ref{subsec:test-space}). If true, \textit{MShape} generates a new input shape $shape_i$ that preserves the same number of dimensions. Otherwise, it randomly generates one shape. Finally, for each layer $layer_i$, \textit{MShape} changes
    the shape of its input to $shape_i$ by padding with a constant value (e.g., pad the shape of $layer_i$ from {\mycode \{10,3,3\}} to {\mycode \{10,5,5\}} with a constant value) or cropping (e.g., crop the shape of $layer_i$ from {\mycode \{10,3,3\}} to {\mycode \{10,1,1\}}).
\end{itemize}

Note that mutating layer inputs may result in improper layer output that cannot be sent to the subsequent layers. Therefore, we design our input-level mutation operators as layer-wise mutations, which means we only target mutating the input of a list of selected layers while not affecting the subsequent layers.
Specifically, when applying \textit{MDtype} to the layer under mutated (denoted as $layer_i$), we first record the original output datatype of $layer_i$, then cast the $layer_i$'s output datatype to the original one after each mutation. Similarly, when applying \textit{MShape} and \textit{MDims}, we also record $layer_i$'s output shape (denoted as $shape^o_i$). After each mutation, we check whether the layer output shape after mutation (denoted as $shape^m_i$) is the same as $shape^o_i$. If not, we reshape $shape^m_i$ to $shape^o_i$. Specifically, we first expand $shape^m_i$'s dimensions or randomly drop some dimensions to make the number of dimensions in $shape^m_i$ consistent with $shape^o_i$. Then, for each dimension, we apply padding with random constant values or cropping to make the total number of elements in each dimension in $shape^m_i$ is consistent with $shape^o_i$.

Three examples on the left of Figure~\ref{fig:example_input_mutation} are three mutants generated by input-level mutations. \textit{MDtype} changes the datatype of {\mycode Conv2D}'s input from {\mycode float32} to {\mycode float16} so we can test precision-specific code in this layer. \textit{MDims} mutates {\mycode BatchNormalization}'s input from a 4D tensor to a 3D tensor; this change can test library codes designed for the tensor with a specific number of dimensions. \textit{MShape} changes the tensor shape of {\mycode Conv2D} from {\mycode (5,8,8,3)} to {\mycode (5,8,5,3)}; thus can test library codes related to the specific tensor shape. After mutating the layer inputs, all three mutation operators will cast or reshape the layer's output to that before mutation.

\subsubsection{\textbf{Parameter-Level Mutation.}}
The design of parameter-level mutation targets mutating the layer parameter values. Similar to input-level mutation, the parameter-level mutation also prioritizes the uncovered mutation option (i.e., parameter value) to diversify layer parameters effectively. Moreover, like input-level mutation, our parameter-level mutation will also randomly choose $n$ layers to mutate. We design our parameter-level mutation, i.e., \textit{MParam}, as follows:

\begin{itemize} 
    \item \textbf{Mutate Parameter (MParam)}: Given a seed model and a list of layers $\{layer_1,..., layer_n\}$, \textit{MParam} searches through all generated models and chooses the parameter value ($value_i$) of a specific parameter ($param_i$) that have not been covered by parameters of $layer_i$. Finally, for each $layer_i$, MParam changes the $param_i$ of it to $value_i$.
\end{itemize}

The example on the right of Figure~\ref{fig:example_input_mutation} is a mutant generated by parameter-level mutation. In this example, parameter-level mutation randomly chooses numeric parameters (e.g., {\mycode strides} and {\mycode dilation\_rate}) and non-numeric parameters (e.g., {\mycode activation} to mutate. By changing the value of {\mycode activation} parameter from {\mycode relu} to {\mycode sigmoid}, this mutant can test new activation functions in the {\mycode Conv2D}. Besides, changing the value of {\mycode strides} and {\mycode dilation\_rate} will influence the padding logic in the {\mycode Conv2D}. Note that though both our \toolname and a prior work~\cite{audee} mutate parameters, prior work randomly mutates parameter values, while \toolname considers the uncovered parameter values. By doing so, \toolname could be more effective in exploring library code.

\subsubsection{\textbf{Structure-Level Mutation}}

\begin{figure}[t!]
    \centering
    \resizebox{0.8\linewidth}{!}{
    \includegraphics{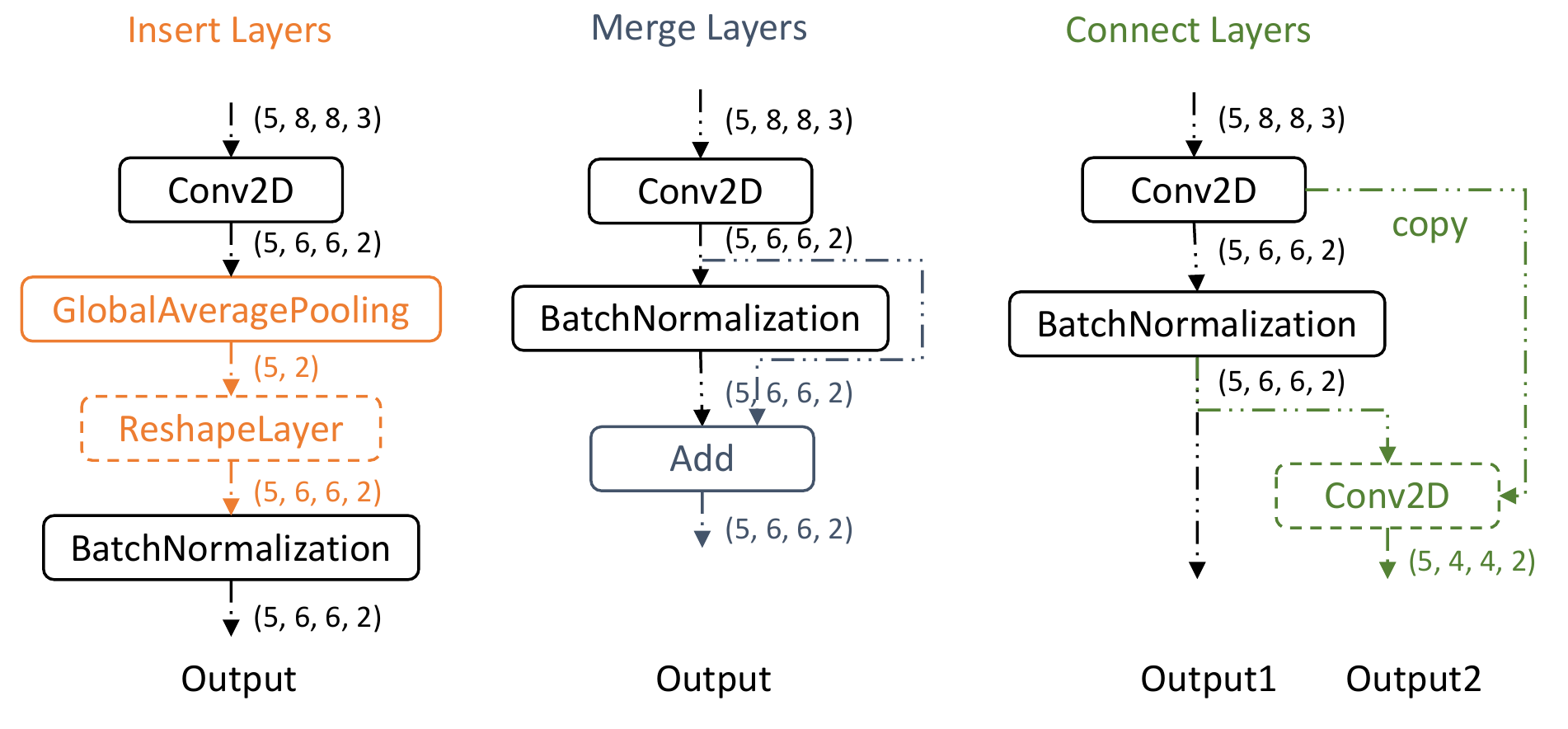}
    }
\caption{Example of Structure-Level Mutation}
    \label{fig:example_structure_mutation}
\end{figure}

The structure-level mutation targets diversifying the structure of a seed model. In particular, it intends to explore the possible layer sequences within the search space. The most straightforward way is to insert some layers into the seed model. However, because of the huge search space when mutating a model's structure, randomly inserting layers~\cite{lemon,deepmutation} is weak in generating an effective mutant that contains new layer sequences. Therefore, our structure-level mutation adopts a similar strategy as input-level and parameter-level mutation. We choose the mutation option that can generate the mutant with new layer sequences in each mutation. Moreover, since some structures cannot be covered by simply inserting layers into the seed model, we also provide some other mutation operators to generate them. Specifically, we design our structure-level mutation as follows:

\begin{itemize} 
    \item \textbf{Insert Layers (IL)}: 
    Given a seed model and the number of layers to be inserted (denoted as $n$), \textit{IL} chooses $n$ layers ($\{layer_1,..., layer_n\}$) from all possible types of layers which previously generated models have not covered. If \textit{IL} cannot find enough uncovered layers, it randomly chooses one instead. For each $layer_i$ ($i\in[0,n]$), \textit{IL} searches for a $layer_i'$ in the seed model so that the layer pair $\left<layer_i',\space layer_i\right>$ not covered by previously generated models can be inserted into this seed model. 
    If no possible $layer_i'$ can be found, \textit{IL} randomly chooses one layer inside the seed model instead. 
    Finally, for each $layer_i$, \textit{IL} inserts $layer_i$ after $layer_i'$ and directs the output of $layer_i$ to the original subsequent layer of $layer_i'$ so the $layer_i$ can be inserted into the seed model. To ensure that the insertion of $layer_i$ follows the dimension constraint and shape constraint required by its subsequent layer, \textit{IL} further applies the same reshape strategy described in \S\ref{alg:input-mutation} to reshape the $layer_i$'s output shape to the original input shape of its subsequent layer.
    
    \item \textbf{Merge Layers (ML)}: 
    Given a seed model, \textit{ML} chooses an uncovered merging layer $layer_m$ which previously generated models have not covered. If previously generated DL models have covered all merging layers, \textit{ML} randomly chooses one instead.
    As described in Figure~\ref{fig:model_constraint}, ML needs to follow the shape constraint, i.e., input tensors of the inserted merging layer should have the same shape. To do so, \textit{ML} scans the seed model and randomly chooses two layers (denoted as $layer_i$ and $layer_i'$) whose tensor output is the same, then merges them using $layer_m$. If no $layer_i$ and $layer_i'$ can be found, \textit{ML} will directly raise an exception, and \toolname will continue the next mutation.
    ML further directs the output of $layer_m$ to the subsequent layer of either $layer_i$ or $layer_i'$ so $layer_m$ can be inserted into the seed model.
    \item \textbf{Connect Layers (CL)}: 
    Given a seed model and the number of layer sequences to be inserted (denoted as $n$), \textit{CL} chooses $n$ layer pairs ($\left<layer_i, layer_i'\right>$, $i\in[0,n]$) inside the seed model that previously generated models have not covered. 
    If \textit{CL} cannot find enough layer pairs, it randomly chooses two layers inside the seed model to form a layer pair instead. 
    For each layer pair $\left<layer_i, layer_i'\right>$, \textit{CL} then connects $layer_i$ and $layer_i'$ by sending the output of $layer_i$ to $layer_i'$. Note that directly connecting $layer_i$ and $layer_i'$ will break the original connection between $layer_i'$ and its preceding layer (denoted as $layer_p$), leaving $layer_p$ detached from the seed model. To avoid this happens, \textit{CL} first makes a copy of $layer_i'$ to keep the original connection between $layer_i'$ and $layer_p$, then it directs the output of $layer_i$ to the copied $layer_i'$. Since the output shape of the copied $layer_i'$ can be different from that of the original $layer_i'$, directing the output of copied $layer_i'$ to the subsequent layers may result in an invalid model that violates shape constraint or dimension constraint. Therefore, \textit{CL} appends the output tensor of the copied $layer_i'$ to the final output lists of the mutant, so the mutant is still valid.
\end{itemize}

Figure~\ref{fig:example_structure_mutation} exemplifies three mutants generated by structure-level mutations. \textit{IL} inserts the \\{\mycode GlobalAveragePooling} after {\mycode Conv2D}. Since the output shape of {\mycode GlobalAveragePooling} is {\mycode (5,2)}, which is inconsistent with the original output shape {\mycode (5,6,6,2)} and cannot be accepted by {\mycode BatchNormalization}, \textit{IL} further reshapes the output shape to the original one. Since the output shape of the {\mycode Conv2D} layer and the {\mycode BatchNormalization} are the same, \textit{ML} can merge these two outputs using the {\mycode Add} layer. \textit{CL} introduces the new layer sequence {\mycode Conv2D} $\rightarrow$ {\mycode BatchNormalization} by connecting these two layers. It further appends the output tensor of the copied {\mycode Conv2D} to the final output lists of the mutant.

\subsubsection{\textbf{Anomaly Mutation}}
Besides diversifying layer inputs, parameter values, or sequences, we noticed that DL libraries lack consideration for some particular values, such as {\mycode NaN}. Indeed, we observed that the manifestation of some bugs requires these special values. Following these observations, we design a \textit{SpecialI} mutation operator to test the layer API using these values. Specifically, given a seed model, \textit{SpecialI} randomly chooses a layer (denoted as $layer_i$) and changes the input value of $layer_i$ to {\mycode NaN} or {\mycode Inf}.

Note that our mutation operators are designed to apply legitimate modification on the seed model to pass validity checks in DL libraries. However, there are still a few invalid mutants generated. For instance, the mutant may require too large GPU memory that directly raises an out-of-memory issue in the DL library; the shape of layer input may be too small, and thus some layers with typical configurations cannot process it. These mutants cannot be successfully constructed by DL libraries. For these invalid models, we follow existing works~\cite{lemon,muffin} to ignore them and continue the next round of mutation.
\subsection{Model Synthesis}
\label{sec:methodology-reduction}
This section proposes a method to synthesize initial seed models based on the collected published DL models (denoted as original models). Our model synthesis method aims to reduce the original model's size while preserving its diversity. Specifically, for an original model, our model synthesis algorithm constructs a new model which covers the layer API calls inside the original model, and this new model is used as the initial seed model for further mutation. However, randomly generating a model does not work well for the following reasons.
On the one hand, randomly building a model from scratch likely generates invalid models, which DL libraries cannot load, and thus cannot be used for testing.
Taking Figure~\ref{fig:model_constraint} as an example, layer {\mycode Add} requires a specific shape constraint (i.e., the shape of two inputs is the same), and layer {\mycode Conv2D} requires the input to have four dimensions. Arbitrary removal of duplicated layers may break such input constraints and unable to generate a valid model. On the other hand, randomly generating DL models cannot effectively achieve the goal of our model synthesis method, i.e., reduce the size of original models while retaining their layer inputs, layer parameters, and layer sequences.

We synthesize a new model using our mutation operators, as introduced in \S\ref{sec:methodology-mutation}. Specifically, we regard the model synthesis as an iterative mutation process using our mutation operators until all layer API calls (i.e., layer inputs, layer parameter values, and layer sequences) in the original model are covered. Given a published model as the original model, we first extract the covered layer types and their parameter values to instantiate a list of candidate layers to be inserted. We further use \textit{IL} operator to insert these layers so the synthesized model can cover all layer APIs and layer parameter values inside the original model. \textit{MDtype}, \textit{MDims}, and \textit{MShape} are further used to retrieve the original model's layer input into the synthesized model. Finally, we use \textit{CL} and \textit{ML} to cover the remaining layer sequences inside the original model.

Algorithm~\ref{alg:reduction} shows the logic of our model synthesis method. Given a published DL model, we first analyze the layer input ($\mathcal{D}_{i}$), parameter ($\mathcal{D}_{p}$), and sequences ($\mathcal{D}_{s}$) covered by the published model. Based on the collected diversities, we then use \textit{IL} operator to insert \textit{layer\_class} with target \textit{layer\_param} to the synthesized model $\mathcal{M'}$ (lines 7-10). In particular, the layer is appended to the end of $\mathcal{M'}$, so we can insert it without breaking any layer sequence diversity. After all $\mathcal{D}_{p}$ are covered, \textit{MDtype}, \textit{MDims}, and \textit{MShape} will be applied to the synthesized model to cover $\mathcal{D}_{i}$ (lines 11-18). Finally, if some layer sequences remain uncovered, the \textit{CL} and \textit{ML} operator is further used to cover these layer sequences (lines 19-22). The model synthesis algorithm stops when the synthesized model covers all layer inputs, layer parameters, and layer sequences inside the original one or the given time budget (i.e., 5 minutes) is reached.

\begin{algorithm}[thbp]
    \caption{Model Synthesis}\label{alg:reduction}
    \LinesNumbered
    \SetKwProg{Def}{def}{:}{}
    \SetKwData{DisI}{$\mathcal{D}_{i} \setminus \mathcal{D}'_{i}$}
    \SetKwData{DisP}{$\mathcal{D}_{p} \setminus \mathcal{D}'_{p}$}
    \SetKwData{DisS}{$\mathcal{D}_{s} \setminus \mathcal{D}'_{s}$}
    \SetKwData{OModel}{$\mathcal{M}$}
    \SetKwData{SModel}{$\mathcal{M'}$}
    \KwIn{
        \OModel: original model
    }
    \KwOut{
        \SModel: synthesized model
    }
    \Def{ModelSynthesis(\OModel)}{
        $\mathcal{D}_{i}$, $\mathcal{D}_{p}$, $\mathcal{D}_{s}$ $\leftarrow$ CollectDiversity(\OModel)\\
        \SModel $\leftarrow \phi$\\
        $\mathcal{D}'_{i}$, $\mathcal{D}'_{p}$, $\mathcal{D}'_{s}$ $\leftarrow$ CollectDiversity(\SModel)\\
        $dis$ $\leftarrow$ \DisI $\cup$ \DisP $\cup$ \DisS\\
        \While(){$dis \neq \phi \land \neg$timeout()}{
        \If{\DisP $\neq \phi$}{
            \textit{layer\_class, layer\_param} $\leftarrow$ rand.Choice(\DisP)\\
            \SModel $\leftarrow$ \textit{IL(layer\_class, layer\_param)}\\
            continue\\
        }
        \If{\DisI $\neq \phi$}{
            \textit{layer\_class, dtype, ndims,shape} $\leftarrow$ rand.Choice(\DisI)\\
            \If{\textit{dtype} $\neq \phi$}{
            \SModel $\leftarrow$ MDtype(\SModel, $\left<\textit{layer\_class}, \textit{dtype}\right>$)\\
            continue\\
            }
            \If{\textit{ndims} $\neq \phi$}{
            \SModel $\leftarrow$ MDims(\SModel, $\left<\textit{layer\_class}, \textit{ndims}\right>$)\\
            continue\\
            }
            \If{\textit{shape} $\neq \phi$}{
            \SModel $\leftarrow$ MShape(\SModel, $\left<\textit{layer\_class}, \textit{shape}\right>$)\\
            continue\\
            }
        }
        \If{\DisS $\neq \phi $}{
            \textit{layer\_sequence} $\leftarrow$ rand.Choice(\DisS)\\
            \SModel $\leftarrow$ \textit{CL(\SModel, layer\_sequence)}\\
            continue\\
        }
        $\mathcal{D}'_{i}$, $\mathcal{D}'_{p}$, $\mathcal{D}'_{s}$ $\leftarrow$ CollectDiversity(\SModel)\\
        $dis$ $\leftarrow$ \DisI $\cup$ \DisP $\cup$ \DisS\\
        }
        \textbf{return} \SModel
    }
\end{algorithm}

\begin{figure}[htbp]
    \centering
    \includegraphics[width=0.8\linewidth]{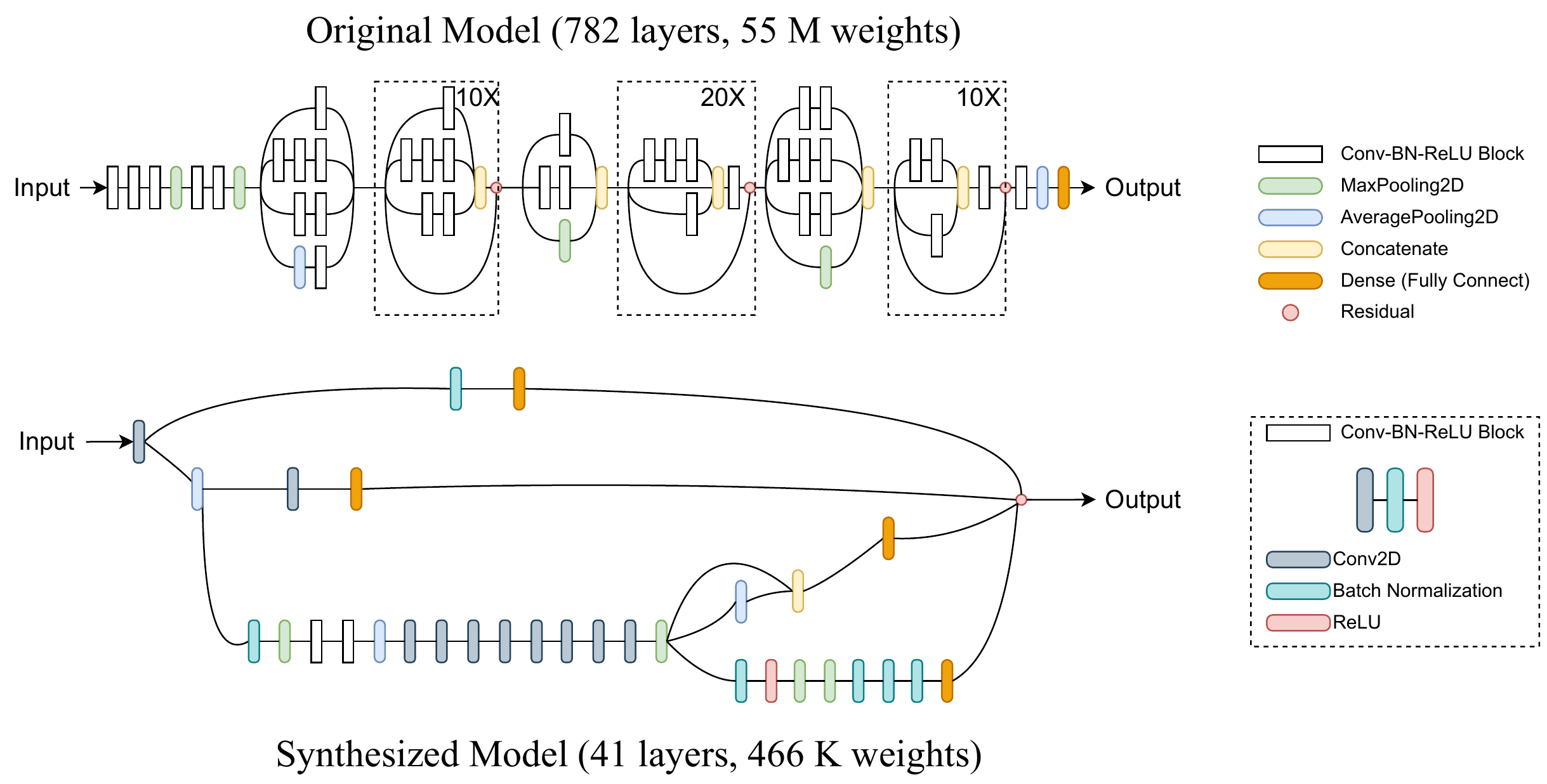}
    \caption{Illustration of Model Synthesis Based on InceptionResNetV2}
    \label{fig:example_architecture_reduction}
\end{figure}

Figure~\ref{fig:example_architecture_reduction} shows a model synthesized from InceptionResNetV2. The original model contains many duplicated layers (e.g., the {\mycode Conv-BN-ReLU block}), which cannot contribute to the diversity of layer API calls. By extracting the layer inputs, layer parameter values, and layer sequences covered by the original model, we can use our model synthesis algorithm to synthesize a more lightweight model while preserving the layer API calls covered by the original one. Note that the synthesized model is generated from scratch and only targets recovering the layer inputs, layer parameter values, and layer sequences covered by the original model. Its topology may be different from the original. As a result, we can reduce the total number of layers in the original model from 782 to 41 (i.e., reduce 94.8\% number of layers) and the total number of weights from 55 million to 466 thousand (i.e., reduce 99.2\% weights). Moreover, the runtime overhead for TensorFlow to load and make inferences on the synthesized model can be lessened from 65 seconds to six seconds while the layer API calls are preserved.

\subsection{Diversity-Driven Model Generation}
\label{sec:methodology-mcmc}
Since not all mutation operators are equally effective, random selection may be ineffective in searching for diverse models in the huge search space. Note that our main objective is to diversify the generated model. Therefore, we should generate DL models for increasing the layer input coverage, layer parameter coverage, and layer sequence coverage. Besides, we need to cover more branches inside the DL library to achieve a higher test coverage. 
Following these objectives, we design our fitness function based on these coverage criteria to measure the potential effectiveness of mutation operators. However, only selecting the mutation operators with the highest fitness score is not optimal since the fitness score is based on the historical iteration and cannot totally represent future results. Intuitively, mutation operators should have a certain possibility to be selected, and those are more likely to diversify DL models and test library code should be more likely to be chosen. This intuition motivates our adoption of the Metropolis-Hasting algorithm, an MCMC (Markov Chain Monte Carlo) sampling method that has been popularly used~\cite{chen2015guided,lemon} for mutation selection. In each iteration of model generation, \toolname prioritizes the mutation-selection effort to prefer those mutation operators that are more effective in increasing our fitness score.

\textbf{Fitness Function}.
Following existing works~\cite{lemon, chen2015guided}, we define the mutation operator ($op$) as the seed in the MCMC sampling process. After selecting a mutation operator $op$, \toolname can generate a new model $m'=op(m)$ by applying the chosen mutation operator $op$ to a randomly selected seed model $m$. 
For each $op$, if applying $op$ is more likely to generate a model that either increases the layer API call diversity or covers new branches inside DL libraries (i.e., increases branch coverage), we should prefer it for future model generation. Therefore, the fitness score of $op$ depends on the quality of $m'$ generated by it. We first introduce how we measure the quality of each generated model $m'$ below:

\begin{align}
    score(m') = \lambda\cdot diverse(m')+(1-\lambda)\cdot branch(m')
    \label{eq:score}
\end{align}

Based on our intuition, we consider two factors when designing $score(m')$: whether it is a diverse model (denoted as $diverse(m')$) and whether it can cover new branches (denoted as $branch(m')$). The higher the $score(m')$ is, the more influential the model $m'$ is in increasing the model diversity and covering new branches in DL libraries. We set $\lambda=\frac{1}{2}$ so that \toolname prefers model diversity and branch coverage equally. To design $diverse(m')$, we compare the layer inputs ($\mathcal{D}'_i$ and $\mathcal{D}_i$), parameters ($\mathcal{D}'_p$ and $\mathcal{D}_p$), and sequences ($\mathcal{D}'_s$ and $\mathcal{D}_s$) covered by $m'$ and $M$ where $M$ is the set of previously generated models. If $m'$ contains new layer types, layer pairs, or layer parameters, $diverse(m')$ will return one, indicating that a diverse model is found; otherwise, $diverse(m')$ returns zero. 
\begin{align}
    diverse(m') &=
    \begin{cases} 
    1, & (\mathcal{D}'_{i} \setminus \mathcal{D}_{i})\cup(\mathcal{D}'_{p} \setminus \mathcal{D}'_{p})\cup(\mathcal{D}_{s} \setminus \mathcal{D}'_{s}) \neq \phi\\
    0, & else
    \end{cases}
    \label{eq:diverse}
\end{align}

To design $branch(m')$, we follow existing strategies~\cite{chen2015guided} to use Equation~\ref{eq:branch}. Suppose the total branches covered by previously generated models is $\mathcal{B}_{M}$ and the branches covered by $m'$ is $\mathcal{B}_{m'}$, if $\mathcal{B}_{m'}$ contains branches that $M$ has not covered, $branch(m')$ will return one; otherwise, $branch(m')$ returns zero.

\begin{align}
    branch(m') &=
    \begin{cases} 
    1, & \mathcal{B}_{M} \setminus \mathcal{B}_{m'}\neq \phi\\
    0, & else
    \end{cases}
    \label{eq:branch}
\end{align}

Finally, given a list of models (i.e., $M_{op}$) generated by mutation operator $op$, we can define the fitness function for $op$ using Equation~\ref{eq:fitness} where $N_{op}$ denotes the total number of models in $M_{op}$.

\begin{align}
    fitness(op) = \frac{\sum^{M_{op}}_{m'}{score(m')}}{N_{op}}
    \label{eq:fitness}
\end{align}

\textbf{Seed Selection}.
\toolname adopts the MH algorithm, a Markov Chain Monte Carlo (MCMC) algorithm commonly used by testing techniques to guide the seed selection~\cite{classfuzz,lemon}.
The MH algorithm is designed for obtaining random samples from a probability distribution. 
It works by generating a sequence of samples whose distribution closely approximates the planned distribution. 
Samples are produced iteratively, where the acceptance of the subsequent sampling (say $s_2$) depends only on the current one (say $s_1$). 
In our setting, a sample corresponds to a seed, i.e., a mutation operator.

Like existing works~\cite{10.1145/2908080.2908095,lemon}, we use the geometric distribution as the desired distribution. 
The geometric distribution is the probability distribution of the number X of Bernoulli trials needed to obtain one success.
Specifically, if the probability of success on each trial is $p$, the likelihood that the $k^{th}$ trial is the first success is given by $Pr(X=k)=(1-p)^{k-1}p$, 
where $p \in \left(0,1 \right)$ is a hyperparameter related to the total number of seeds~\cite{10.1145/2908080.2908095}. 
We sort all seeds based on their fitness scores computed by Equation~\ref{eq:fitness} from the highest to the lowest. 
By mapping the sorted seeds into the geometric distribution, the sampling process can prefer those with high fitness scores to those with low fitness scores.  Specifically, suppose the last selected seed is $s_1$, the probability of accepting seed $s_2$ is calculated by Equation~\ref{eq:geometric_distribution}. 
In particular, if the $k_2$ (the index of $s_2$ after we sort all seeds) is smaller than $k_1$ (the index of $s_1$), i.e., the fitness score of $s_2$ is larger than that of $s_1$, $P(s_1 \rightarrow s_2) = 1$ and we directly accept $s_2$. 
Otherwise, we accept $s_2$ with a certain probability $(1-p)^{k_2-k_1}$.

\begin{align}
    P(s_1 \rightarrow s_2) = min\left(1, \frac{Pr(s_2)}{Pr(s_1)}\right)=min\left(1, (1-p)^{k_2-k_1}\right)
    \label{eq:geometric_distribution}
\end{align}

Table~\ref{tab:seed_selection_example} is an example demonstrating the seed selection, it lists each seed's fitness score (row 3) in one iteration. Based on these fitness scores, \toolname first sorts all seeds from the highest fitness score to the lowest (row 4), and then it assigns the index $k$ for each seed (row 5) from 1 to 8.
The selection of the next seed $s_2$ relies on the hyperparameter $p$ and the index of the last chosen seed $s_1$. Specifically, \toolname randomly selects a seed $s_2$ and accepts this selection based on acceptance probability (i.e., $P(s_1 \rightarrow s_2)$) calculated by Equation~\ref{eq:geometric_distribution} (in this example, $p=0.4$, and $s_1$ is \textit{MDtype} with the index to be 3).
In particular, when \toolname selects the seed whose index is smaller than or equal to 3 (i.e., \textit{MParam}, \textit{ML}, or \textit{MDtype}), the acceptance probability is 1, which means that \toolname will directly accept this selection. If \toolname selects the seed whose index is larger than 3 (i.e., \textit{IL}, \textit{MDims}, \textit{MShape}, \textit{CL}, or \textit{Special}), \toolname will accept this selection based on the acceptance probability. The higher the selection's acceptance probability is, the more likely \toolname will accept this selection. 
In this example, among seeds whose indexes are larger than 3, selecting \textit{IL} is most likely to be accepted, while selecting \textit{SpecialI} is least likely to be accepted.
\toolname repeats this seed selection until a selection is accepted.

\input{Table-Seed-Selection-Example.tex}
\textbf{Model Pool Update}. As illustrated earlier, in each iteration, \toolname randomly selects a model from the model pool to generate a mutant. In the beginning, the model pool contains all the initial synthesized models. During the model generation, \toolname adds mutants, which can introduce new layer inputs, parameters, or sequences into the model pool to increase the diversity of the seed model. If the total number of seed models in the model pool is larger than a pre-defined size (i.e., 50 following existing work~\cite{lemon}), \toolname will randomly remove a seed model from the model pool.

Algorithm~\ref{alg:sampling} shows the detailed process for \toolname to generate models. 
In each iteration, \toolname uses the MH algorithm to select an operator $op_i$ (in function \textit{SeedSelection}) and randomly select a seed model $m_i$ from the model pool (line 4). A mutant $m'_i$ is thereby generated by applying $op_i$ to $m_i$ (line 5). If the mutant $m'_i$ can cover new layer API calls or cover new branches inside the DL library, \toolname updates the mutant to the model pool $M$ (lines 6-7). Otherwise, \toolname continues the next iteration. The diverse model generation algorithm ends when the given time budget is reached.

\input{Alg-MCMC}

\subsection{Bug Detection}
\label{subsec:bug_detection}

We further use the generated models as test inputs for DL library testing in each iteration. For bug detection, we follow existing works~\cite{cradle,audee,lemon,graphfuzz,muffin} to adopt differential testing as the test oracle. Overall, \toolname targets three types of bugs: \textit{crash}, \textit{Not-A-Number (NaN)}, and \textit{inconsistency bugs}. For crash bugs and NaN bugs, we consider a bug detected if some DL libraries crash or output NaN value when loading and executing the model while some other libraries do not. Inconsistency bug refers to the difference between the model execution results of different libraries. Since a DL model initialized with the same weights should have the same or close model prediction result on different DL libraries, inconsistent prediction results may indicate the inconsistent algorithm implementation of different DL libraries, which is likely to be a bug~\cite{cradle,audee,lemon}. However, because of the computation randomness during the DL library, two libraries' outputs on the same model and the same input are not the same. Therefore, we adopt a well-defined inconsistency metric, i.e., $D\_MAD$, proposed by CRADLE~\cite{cradle} and has been used by existing DL library testing works~\cite{lemon,cradle} to measure the inconsistency degree across different DL libraries for inconsistency bug detection. 
In particular, $D\_MAD$ calculates the inconsistency distance by comparing the relative distance of model outputs from the ground-truth labels~\cite{cradle}. 
Specifically, given two predicted vectors $Y$, $Y'$, and the ground truth $O$, $\delta_{Y,O}$ and $\delta_{Y',O}$ are calculated to measure the distance between the predicted vectors and the ground truth using Equation~\ref{eq:d_mad_delta}, and the distance between two predicted vectors are then measured by Equation~\ref{eq:d_mad_distance}. 
An inconsistency bug is detected when the $D\_MAD$ score is larger than a pre-defined threshold $T$. Note that DL models may have multiple outputs (as exemplified on the right of Figure~\ref{fig:example_structure_mutation}), which is inapplicable to the $D\_MAD$ metrics. Therefore, for DL model with multiple outputs, we reshape its outputs to the same shape and add them together so we can still use $D\_MAD$ for inconsistency bug detection.

\begin{equation}
    \delta_{Y,O}=\frac{1}{N}\sum^N_{i=1} |Y_i-O_i|
    \label{eq:d_mad_delta}
\end{equation}
\begin{equation}
    D\_MAD_{O,Y,Y'}=\frac{|\delta_{Y,O}-\delta_{Y',O}|}{\delta_{Y,O}+\delta_{Y',O}}
    \label{eq:d_mad_distance}
\end{equation}

%% file: Table-Diversity-Space.tex
\begin{table}[htbp]
\caption{Components In Layer API Calls}
\renewcommand{\arraystretch}{0.9}  
\setlength{\tabcolsep}{2.2pt}
\resizebox{0.8\linewidth}{!}{  
\begin{tabular}{l|l|l}
Property   &   Element  &   Space \\ \hline \hline
\multirow{5}{*}{Layer Input}
&   Datatype  & 
\begin{tabular}[c]{@{}l@{}}
Layer A/B/...: \\
    \quad int16 | int32 | float16 | ...\\
\end{tabular}    \\ \cline{2-3}
&   Shape   &
\begin{tabular}[c]{@{}l@{}}
Layer A/B/...: \\
    \quad shape 1 | shape 2 | shape 3 | ...\\
\end{tabular}   \\ \cline{2-3}
&   Number of Dimensions &
\begin{tabular}[c]{@{}l@{}} 
Layer A/B/...: \\
    \quad 2D | 3D | 4D | 5D | ...\\
\end{tabular}    \\ \hline
Layer Parameter   &   Parameter Value   & 
\begin{tabular}[c]{@{}l@{}}
Layer A: \\
    \quad Param A1: v1 | v2 | ...\\ 
    \quad Param A2: v1 | v2 | ...\\
    \quad ... \\
Layer B: \\
    \quad Param B1: v1 | v2 | ...\\
    \quad Param B2: v1 | v2 | ...\\
    \quad ... \\
...
\end{tabular}    \\ \hline
Layer Sequence    &   Pair of Layers    & 
\begin{tabular}[c]{@{}l@{}}
Layer A $\rightarrow$ Layer B\\
Layer B $\rightarrow$ Layer A\\
Layer A $\rightarrow$ Layer A\\
...
\end{tabular} \\ \hline
\end{tabular}
}
\label{tab:model_diversity_space}
\end{table}

%% file: Table-New-Mutation-Operators.tex
\begin{table}[t!]
\caption{Mutation Operators}
\renewcommand\arraystretch{1.6}
\resizebox{1.0\linewidth}{!}{
\begin{tabular}{l|l|l|l}
\hline
Type    &   Names  &  Input   &   Description   \\ \hline \hline
\multirow{3}{*}{Input-Level Mutation}
& Mutate Datatype (MDtype)    &   a model; a list of layers  & Change the datatype of input for layers in the list. \\ \cline{2-4}
& Mutate Dimension (MDims)    &   a model; a list of layers  & Change the number of input dimensions for layers in the list. \\ \cline{2-4}
& Mutate Shape (MShape)    &   a model; a list of layers  & Change the shape of input for layers in the list. \\ \cline{1-4}
\multirow{1}{*}{Parameter-Level Mutation}
& Mutate Parameter (MParam) &  a model; a list of layers  &   Change the value of layer parameters for layers in the list.   \\ \cline{1-4}
\multirow{3}{*}{Structure-Level Mutation}
& Insert Layers (IL)  &   a model; a list of layers   &   Insert a list of layers into model.    \\ \cline{2-4}
& Merge Layers (ML)   &   a model; a merging layer    &   merge two layers' output using the selected merging layer.    \\ \cline{2-4}
& Connect Layers (CL) & a model; a list of layer pairs    &   connect layers according to the layer pair in the list  \\ \cline{1-4}
Anomaly Mutation    & Special Input (SpecialI) & a model; a layer & Change the input of a selected layer to infinity or NaN \\ \cline{1-4}
\end{tabular}
}
\label{tab:method_new_mutation}
\end{table}

%% file: Table-Seed-Selection-Example.tex
\begin{table}[htbp!]
\caption{Example of Seed Selection}
\small
\label{tab:seed_selection_example}
\renewcommand{\arraystretch}{1.0}
\setlength{\tabcolsep}{2.2pt}
\begin{tabular}{l|l|l|l|l|l|l|l|l}
\hline
\multicolumn{9}{c}{Unsorted Seeds}\\ \hline

& MDtype (s1)
& MDims
& MShape
& MParam
& IL
& ML
& CL
& SpecialI\\    \hline
Fitness
& 0.60
& 0.53
& 0.44
& 0.68
& 0.57
& 0.62
& 0.42
& 0.38\\   \hline \hline
\multicolumn{9}{c}{Sorted Seeds}\\ \hline
& MParam
& ML
& MDtype (s1)
& IL
& MDims
& MShape
& CL
& SpecialI\\    \hline
k
& 1
& 2
& 3
& 4
& 5
& 6
& 7
& 8\\   \hline \hline
$P(s_1 \rightarrow s_2)$
& 1
& 1
& 1
& 0.6
& 0.36
& 0.22
& 0.13
& 0.08\\   \hline \hline
\end{tabular}
\end{table}

%% file: Alg-MCMC.tex
\begin{algorithm}[thbp]
    \caption{Diverse Model Generation Algorithm}\label{alg:sampling}
    \LinesNumbered
    \KwIn{
        $\mathbf{M}$: a model pool; $\mathbf{Op}$: a mutation operator pool
    }
    \KwOut{$\mathbf{M'}$: All generated models}
    \SetKwProg{Def}{def}{:}{}
    \Def{ModelGeneration($P$, $M$, $Op$)}{
        \While(){$\neg$timeout()}{
            ${op}_i$ $\leftarrow$ \textit{SeedSelection} ($S$)\\
            $m_i$ $\leftarrow$ rand.Choice($M$)  // randomly choose $m$\\
            $m'_i \leftarrow op_i(m)$\\
            \If{score($m'_i$) > 0}{
                $M \leftarrow$ \textit{UpdateModelPool($M$, $m'_i$)}\\
            }
        }
        \textbf{return} $M'$\\
    }
    \Def{SeedSelection($S$)}{
        // select candidate based on MH algorithm\\
        $S_{sorted} \leftarrow$ \textit{SortSeed($S$)}  // Using Fitness Function (Eq.~\ref{eq:fitness})\\
        $prob \leftarrow 0$\\
        $k_1 \leftarrow$ get the index of last selected seed\\
        \While(){\textit{New Rand()} $\geq prob$ }{
            $s_i \leftarrow$ random.Choice($S_{sorted}$)\\
            $k_2 \leftarrow S_{sorted}.index(s_i)$\\
            $prob \leftarrow (1-p)^{k_2-k_1}$\\
        }
    \textbf{return} $s_i$
    }
    \Def{UpdateModelPool($M$, $m'_i$)}{
        \If{sizeof($M$)>=\textit{pool\_size}}{
            $idx \leftarrow$ rand.randInt(sizeof($M$))\\
            $M'$.pop(idx)\\
        }
        $M'$.add($m'_i$)\\
        \textbf{return} $M'$
    }
\end{algorithm}

%% file: 6-Evaluation.tex
\section{Evaluation}
We evaluated the performance of \toolname from two perspectives: the effectiveness of diverse model generation and bug detection. In particular, we studied four research questions (RQs):
\begin{itemize}[leftmargin=*]
    \item \textbf{RQ1: Can \toolname outperform existing techniques?}
    We first demonstrate the advantage of our diversity-driven approach by comparing \toolname against existing techniques in terms of our proposed coverage criteria (i.e., layer input coverage, layer parameter coverage, and layer sequence coverage) and their branch coverage, line coverage on a representative DL library.
    \item \textbf{RQ2: Can \toolname detect real bugs?}
    To further demonstrate the usefulness of \toolname, we present its bug-revealing capability. We evaluate \toolname on the latest release of eight popular DL libraries: TensorFlow, PyTorch, MXNet, ONNXRuntime, Keras-MXNet, TF2ONNX, and ONNX2PyTorch.
    \item \textbf{RQ3: Can test efficiency be boosted by the synthesized initial models?} 
    We also conducted the experiment to show the impact of the synthesized initial models on the test efficiency by illustrating the trends of coverage growth achieved with and without model synthesis.
    
    \item \textbf{RQ4: To what extent do the proposed diversity-driven mutation operators and the search algorithm contribute to the performance of \toolname?}
    Finally, to further dissect the secrets of \toolname, we conducted an ablation study to analyze two key components (i.e., diversity-driven model mutation operators and the MCMC-based search algorithm) in \toolname, examining their usefulness in generating diverse models and exercising various computations.
    
\end{itemize}

\subsection{Experiment Setup}
\label{subsec:exp_setup}
All the experiments were conducted on a machine powered by Intel Core i9 with one Titan RTX and two 2080Ti GPU cards.

\textbf{Baselines}. 
We considered three state-of-the-art DL testing techniques as baselines in our experiment: CRADLE~\cite{cradle}, LEMON~\cite{lemon}, and Muffin~\cite{muffin}. Since no executables or source codes of GraphFuzz~\cite{graphfuzz} and Audee~\cite{audee} are available, we did not include them as baselines.
We consider the layer input coverage, layer parameter coverage, layer sequence coverage, branch coverage, and line coverage as the metrics to evaluate their overall performance. In particular, the first three coverage criteria are used to measure the layer API call diversity; the branch and line coverage are used to measure their test coverage on DL libraries. The three baselines are designed in different ways. In particular, CRADLE focuses on detecting inconsistency issues exposed by collected DL models and datasets. Since it does not involve the model generation, we collect the total layer inputs, layer parameters, and layer sequences in the initial models. We use our target DL libraries to load their generated models and make inferences on their collected datasets to collect their branch and line coverage.
For LEMON and Muffin, we used their default settings to generate models. Also, since other baselines do not consider the training phase during testing, for Muffin, which targets both training and inference phases, we only considered the inference phase for the fairness of the comparison. 

\textbf{Benchmarks and Implementations}.
\input{Table-Origin-Model-Statistics.tex}
We selected 10 popular DL models with different structures as the original models.
They cover diverse application domains across both image data and sequential data. The three baselines have also used these models in the evaluation.
In particular, eight DL models (i.e., LeNet, AlexNet, DenseNet121, InceptionResNetV2, InceptionV3, MobileNetV2, ResNet50, Xception) are image classification models trained on MNIST~\cite{mnist}, CIFAR-10~\cite{cifar10}, and ImageNet~\cite{imagenet}. 
The remaining two models are LSTM-based models trained on Sine-Wave dataset and Stock-Price dataset by existing work~\cite{lemon}. 
All DL models are online-available on our project site.\footnote{\url{https://github.com/maybeLee/COMET}
} The number of layers, weights, and the diversity we collected for each original model are listed in Table~\ref{tab:origin_model_statistics}.
The setting of hyperparameter $p$ in our algorithm depends on the size of seeds in the MH algorithm~\cite{10.1145/2908080.2908095}. Since the total size of initial seeds (i.e., mutation operators) is 8, following the setting recommended by existing work~\cite{10.1145/2908080.2908095}, the range of $p$ should be set within [0.313, 0.598].
In our implementation, we set $p$ to 0.4 without fine-tuning in our experiment. For inconsistency detection, we set $T$ to 0.4, which is the value used by an existing technique~\cite{lemon}.

\textbf{Layer API call Diversity}.
We also use Keras~\cite{keras} to generate DL models for a fair comparison with LEMON~\cite{lemon} and Muffin~\cite{muffin}. Following existing works~\cite{muffin,audee}, we take the Keras documentation\footnote{https://keras.io/api/layers/} as the reference to define the mutation space. In particular, we collect the information of all 59 Keras layer APIs (i.e., $N_{layer}=59$), including their possible datatypes, required input dimensions, and possible parameter values. Specifically, we set all possible datatypes for each layer API to be {\mycode {`bfloat16', `double', `float16', `float32', `float64', `half'}} (i.e., $n_{type}(l)=6, \forall l \in Layer List$). 
We further manually collect the possible parameter values, possible number of input dimensions, and possible number of output dimensions for each layer API. We set $n_{shape}$ to be 5 to measure the coverage of layer input shape. Following \S\ref{subsec:test-space}, the total number of possible layer sequences is 2170. Since the total number of layer APIs is few and the documentation has identified the possible numbers of input dimensions and possible parameter values, the manual effort in collecting these spaces is trivial.

Like GraphFuzz~\cite{graphfuzz}, when implementing our mutation operators, we inevitably introduce some utility operators into the generated models. Specifically, we add the {\mycode tf.cast} operator into the model to cast the original input's datatype to our required one; we use {\mycode padding}, {\mycode cropping}, and {\mycode reshaping} operators to diversify layer input shape and input dimension. Since models generated by other works do not have these operators, we do not consider them when measuring the diversity of our generated models.

\textbf{DL Libraries Under Test}.
We evaluated the bug detection performance of \toolname on the latest release of eight popular DL libraries: Keras 2.8.0, TensorFlow 2.9.0, PyTorch 1.10.0, MXNet 1.9.0, ONNXRuntime 1.10.0, Keras-MXNet 2.2.4.3, TF2ONNX 1.9.3, and ONNX2Pytorch 0.4.1. Following earlier practice~\cite{cradle, lemon, muffin}, \toolname leverages Keras APIs to generate models uniformly for testing different DL libraries. The generated models are converted for MXNet, ONNX, and PyTorch using three libraries, Keras-MXNet, TF2ONNX, and ONNX2Pytorch. The setup helps to control the variation of models used for different libraries.

\textbf{Unique Bugs Identification}.
Following the practice of existing works~\cite{audee,cradle,lemon}, we manually analyzed each detected bug and identified the unique ones.
In particular, 
we consider a crash bug unique if it has a distinct stack trace. 
We manually located the layer that triggered the bugs for NaN and inconsistency bugs and analyzed the root cause.
If the exact root cause causes multiple bugs, we consider them as duplicated bugs and only consider one of them in the subsequent analysis.

\subsection{RQ1: Comparison with Baselines}
\label{subsec:rq1}
This section compares \toolname with three baselines: CRADLE~\cite{cradle}, LEMON~\cite{lemon}, and Muffin~\cite{muffin}. Specifically, we evaluated their performance by running them on the same 10 published DL models (see \S\ref{subsec:exp_setup}) for six hours. Note that CRADLE tests DL libraries by directly collecting publicly-available DL models. We, therefore, referred to the coverage results of initial DL models to capture CRADLE's performance. Since Muffin does not require initial DL models for model generation, we ran it for six hours using its default setting to capture the coverage result. Although Muffin does not require these DL models for model generation, we can easily obtain these 10 publicly-available DL models used in our experiment online without additional training. Therefore, we argue that using these initial models will not affect fairness when we compare Muffin with \toolname and LEMON.

Table~\ref{tab:evaluation_baseline} presents the coverage results achieved by the baselines. 
Since collecting DL library coverage is costly and testing practices and bugs inside different DL libraries share a significant commonality~\cite{jia2021unittestquality, oracleapproximation, empirical_2022_dlframeworks, chen2022toward}, we used the branch and line coverage on TensorFlow’s model construction and model execution modules as the representative for evaluating each technique’s test coverage.
According to Table~\ref{tab:evaluation_baseline}, \toolname outperforms the baselines on all five coverage criteria. In particular, the layer input coverage achieved by \toolname (69.7\%) is 35.6\% higher than the best result among the baselines (34.1\%); and the layer parameter coverage achieved by \toolname (50.2\%) is 24.3\% higher than the best result achieved by the baseline (25.9\%). \toolname can also cover over twice as many layer sequences as the state-of-the-art (39.0\% vs 15.6\%).

Besides the improvement of layer API call diversity, the results suggest a correlation between the test coverage and the diversity of the generated models. Specifically, \toolname outperforms the baselines by covering at least 3.4\% more branches and 4.5\% more lines in TensorFlow's model construction and model execution modules. Nevertheless, the issue of relatively low branch coverage is common to fuzzing-based techniques. We further performed an in-depth analysis towards the uncovered covered code and discussed it in \S\ref{sec:discuss_uncovered}.

Overall, \toolname outperforms the state-of-the-art approaches. In particular, compared with the optimum performance achieved by baselines, \toolname improves 35.6\%, 24.3\%, and 23.4\% on layer input coverage, layer parameter coverage, and layer sequence coverage. In addition, \toolname can outperform the baselines by covering at least 3.4\% more branches and 4.5\% more lines in TensorFlow's model construction and model execution modules.

\input{Table-Evaluation-Baseline}

\subsection{RQ2: Effectiveness of Bug Detection}
\input{Table-Evaluation-Bug}

Table~\ref{tab:evaluation-bug} demonstrates the new bugs detected by \toolname across various DL libraries, categorized by their bug symptoms.
In total, \toolname detected \detectedbugs new bugs in the latest release version of evaluated DL libraries. After we reported them, \confirmedbydevelops of them have been confirmed by DL library developers and \fixedbugs out of those confirmed bugs were fixed. The remaining 11 bugs have been reported and are waiting for developers' confirmation. 
The detected bugs can be categorized according to three different symptoms. We further conduct an in-depth analysis of these bugs to answer two questions: 1) Are diversifying layer API calls practical in generating DL models to detect bugs? 2) Are there any other diversities that can help detect bugs? The first question examines the usefulness of our insight; if the manifestation of the majority of bugs requires a specific layer API call (i.e., a particular layer input, parameter value, or sequence), diversifying layer API calls can help detect DL library bugs. The second question aims to evaluate the completeness of our diversities. If the majority of the bugs are detected by other diversities, diversifying layer inputs, layer parameters, and layer sequences may be insufficient.

\textbf{Crash Bugs}. Among the bugs detected, 23 are crash bugs. Developers have confirmed 13 bugs and fixed five out of these confirmed bugs.
We further analyzed the 13 confirmed crash bugs. Four occurred during model conversion (three in TF2ONNX library and one in ONNX2PyTorch library). The remaining eight crash bugs occurred during model execution (two in ONNXRuntime and six in Keras).
We noticed that \textbf{only one crash bug is caused by a single layer with its default parameter values}.
11 of the 13 crash bugs were manifested by either a specific layer input, a specific layer parameter, or a specific layer sequence. For instance, sending a tensor with {\mycode bfloat16} datatype to {\mycode Conv2D} will cause Keras crashes; failing to set either {\mycode strides} or {\mycode dilation\_} {\mycode rate} in SeparableConv2D to 1 can result in the wrong shape inference, causing Keras to crash; layer sequence between two {\mycode ThresholdReLU} layers will cause TF2ONNX crashes.
\textbf{The experiment result affirms the importance of generating models with diverse layer inputs, parameters, and sequences.} 
Moreover, \toolname detected two severe crash bugs that directly led to core dump issues in Keras and ONNXRuntime. Among these two bugs, one is caused by a particular layer sequence (i.e., Dense-Dot) and a particular layer parameter (i.e., {\mycode units}=0); the other one is caused by a specific layer input shape (i.e., tensor with an empty shape for {\mycode Conv3DTranspose} layer).
For these two issues, developers instantly created pull requests to fix them after our reporting.

\textbf{NaN Bugs}. 
Besides crash bugs, we also detected eight Not-A-Number (NaN) bugs, seven out of which have been confirmed by developers, including two fixed ones. We further investigated the fault-triggering conditions for these confirmed NaN bugs. We found
that four out of seven NaN bugs are caused by a particular layer API call, while the remaining three NaN bugs are caused by a particular value: NaN. Specifically, two confirmed NaN bugs were manifested by particular layer parameters, such as setting the {\mycode max\_value} in the {\mycode ReLU} layer to a concrete value instead of ``None'' (the default option). One NaN bug was caused by a specific layer pair (i.e., Conv2D+Multiply). One NaN bug was caused by a specific layer input shape (i.e., setting the shape of {\mycode ConvLSTM2D} input's first dimension to ``1''). We noticed that the remaining three NaN bugs were exposed when we passed a NaN value to a layer with its default parameter. Although the manifestation of these three NaN bugs does not require a specific layer input, layer parameter, or layer sequence, it still requires a special value: NaN, indicating the usefulness of our mutation operator \textit{SpecialI} in detecting bugs.

\textbf{Inconsistency Bugs}.
One confirmed bug is the inconsistency bug between the TensorFlow library and the ONNXRuntime library. We find that this bug lies in the model conversion phase in TensorFlow. Specifically, when TensorFlow converts the model from the Keras format to TensorFlow Protobuf format (a private format of DL model inside the TensorFlow library), the model will trigger the bug if the {\mycode axis} parameter in the Softmax layer is set to ``-2''. 

\input{Table-Discussion-Model-Chracteristics}

To better understand the characteristics of DL library bugs detected by us, we summarized the fault-triggering condition of our confirmed bugs. As shown in Table~\ref{tab:discussion_memo_bugs}, most bugs (i.e., 18 out of \confirmedbydevelops) detected require a specific layer input, layer parameter value, or layer sequences. In contrast, only one bug is manifested by a single layer API with its default parameter. Therefore, only covering the layer API is not enough to detect DL library bugs.

In conclusion, our experiment shows that \toolname effectively detects bugs in DL libraries. More importantly, we found that 18 out of the \confirmedbydevelops bugs confirmed by developers can only be triggered by deliberate layer inputs, layer parameters, and layer sequences. The remaining three bugs are manifested by directly calling a layer API with a particular value: NaN. 
The finding further supports the insight of \toolname, i.e., models with diverse layer API calls help detect DL library bugs.

\subsection{RQ3: Performance of Model Synthesis}
\label{eval_simplification}
In the above RQs, we evaluate the effectiveness and significance of \toolname by assessing the diversity of the generated models, the achieved test coverage, and the real new bugs detected by \toolname. In the following, we further investigate the effectiveness of the major components in \toolname. Specifically, we aim to examine the effectiveness of model synthesis in this RQ by answering the following two questions: 1) Can our model synthesis method successfully reduce the size of original models while preserving their layer API call diversity? 2) Can our model synthesis method boost the efficiency of DL library testing?

\textbf{Model Synthesis Results}.
We first study the effectiveness of our model synthesis method. Specifically, we investigate the results of our model synthesis method, including the size of synthesized models and the diversity of synthesized models. To comprehensively evaluate the synthesis results, we apply our model synthesis method to the original DL models 20 times and present the results in Table~\ref{tab:evaluation_synthesis}. 
As shown in Table~\ref{tab:evaluation_synthesis}, the 10 published models (denoted as original models) used by existing techniques contain 2025 layers and over 300 million weights. However, these original models can only cover 116 unique layer inputs, 122 unique layer parameter values, and 25 unique layer sequences. Compared with these original models, models synthesized by our method contain an average of 318 layers in total. At the same time, they can cover 123.75 layer inputs, 153.85 layer parameters, and 45.25 layer sequences on average, including all layer inputs, parameters, and sequences covered by the original models. Moreover, our synthesized models can provide an even higher diversity of layer parameters and layer sequences. Such results reflect that our model synthesis method can not only significantly reduce the original models' size, but also can preserve and even increase their diversities.

\begin{figure}[t!]
    \centering
    \includegraphics[width=1.0\linewidth]{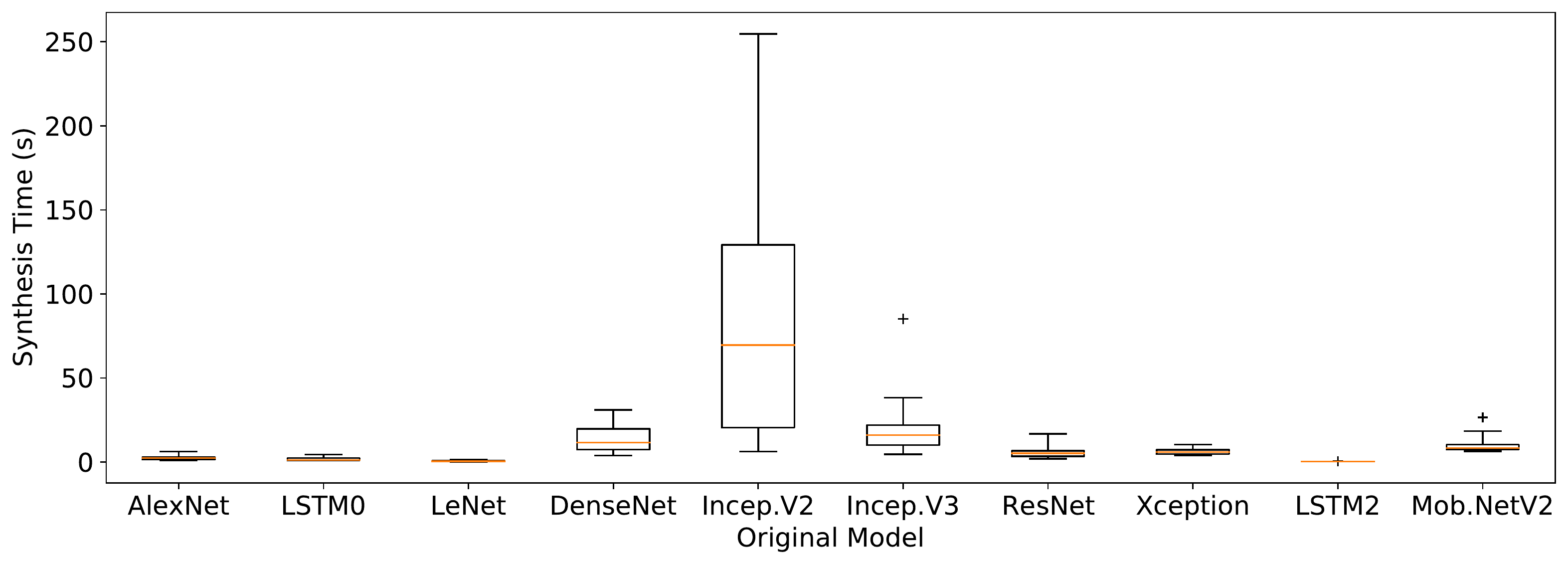}
    \caption{Model Synthesis Time For Each Original Model}
    \label{fig:rq3_synthesis_efficiency}
\end{figure}

We further investigate the efficiency of our synthesis method. As is shown in Figure~\ref{fig:rq3_synthesis_efficiency}, for most original models, our model synthesis method can synthesize a new one within one minute (the average synthesis time for most DL models is less than 20 seconds). We notice that the synthesis time of the InceptionResNetV2 model is comparably larger than the others. The reason is that InceptionResNetV2 is carefully crafted with more complex structures (i.e., it contains 782 layers). It is not easy to analyze and generate a valid DL model that can cover all layer inputs, layer parameters, and layer sequences inside it. Therefore, our model synthesis algorithm needs to spend more time searching until a valid model is found. Nevertheless, since the model synthesis will only be applied once to the original model and the maximum synthesis time for each model is 4 minutes, our model synthesis method is feasible and affordable. 

\input{Table-Evaluation-Reduction}

\textbf{Comparison of Efficiency Between Synthesized Models and Original Models}.
To investigate how our synthesized models can boost the efficiency of our diverse model generation, we prepared two sets of initial seed models based on the original DL models collected in \S\ref{subsec:exp_setup} and the models synthesized from these original models. We run our diverse model generation algorithm on each set of initial seed models for six hours. We observed a significant improvement in terms of efficiency. In particular, when we used the original models as initial seeds, the average model generation time overhead is 38.10 seconds. In contrast, the average model generation time overhead is reduced to 24.35 seconds when we used the synthesized models as initial seeds. As a result, under the same time budget, using synthesized models as initial seeds can generate 888 models, 56.33\% larger than the total number of models generated when using the original models (i.e., 568).

\begin{figure}[t!]
    \centering
    \includegraphics[width=1.0\linewidth]{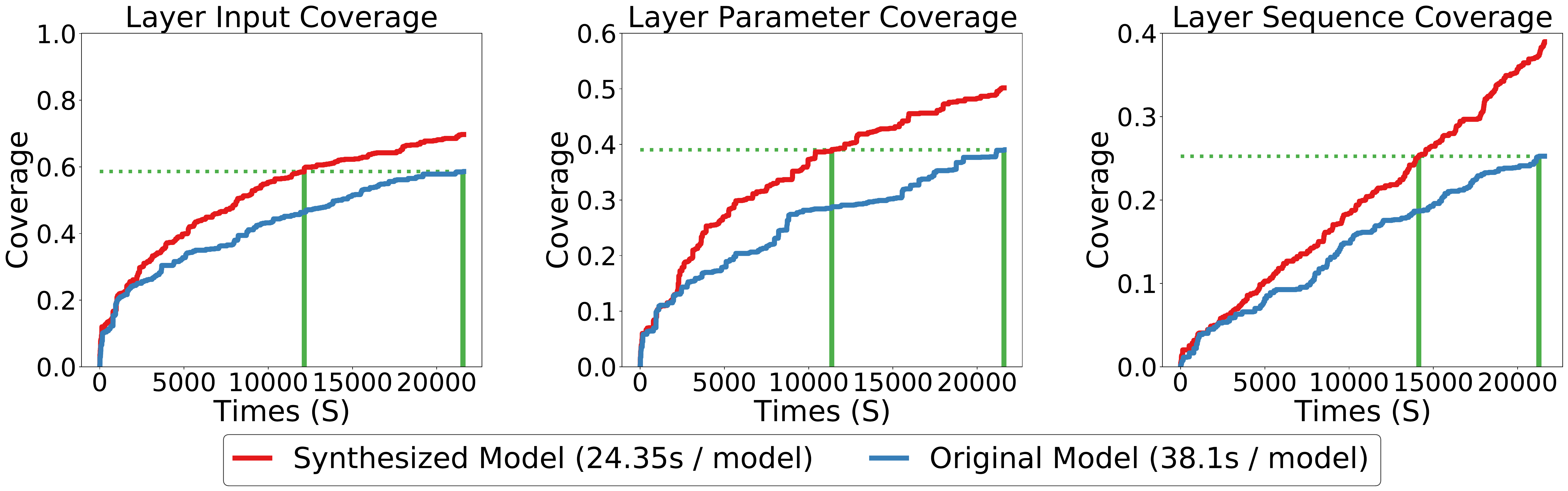}
    \caption{Coverage by using Original Models and Synthesized Model}
    \label{fig:rq3_synthesis_effectiveness}
\end{figure}

We further record the layer input coverage, layer parameter coverage, and layer sequence coverage during the model generation. Figure~\ref{fig:rq3_synthesis_effectiveness} presents the coverage results when using the synthesized models as the initial seed model (in red) or the original models as the initial seed model (in blue). 
The coverage result demonstrates a clear improvement after using the synthesized models as the initial seed models. Specifically, the layer input coverage rises from 58.6\% to 69.7\%, layer parameter coverage increases from 39.0\% to 50.2\%, and layer sequence coverage improves from 25.3\% to 39.0\%.

In summary, compared with the original models, our model synthesis method can efficiently synthesize models with much smaller sizes while retaining the original models' diversity. Using these synthesized models as the initial seed models for further model generation, we can significantly improve the layer input coverage, layer parameter coverage, and layer sequence coverage compared with those using the original models as the initial seed models. 

\subsection{RQ4: Ablation Study on Mutation Operators and Search Algorithm}

We further evaluate the effectiveness of our diversity-driven mutation operators and MCMC-based search algorithm. The diversity-driven mutation operators aim to target mutating with diverse layer API calls. The MCMC-based search algorithm is designed to guide model generation towards a higher layer API call diversity on DL libraries. 

To conduct an ablation study of the proposed mutation operators, we constructed another baseline named \memoo. 
Instead of using our proposed mutation operators, \memoo adopts the existing mutation operators designed by LEMON~\cite{lemon}, DeepMutation~\cite{deepmutation}, and Audee~\cite{audee}. These mutation operators are designed to randomly mutate the model's architecture and parameters. If \toolname can outperform \memoo, our proposed mutation operators would be more helpful in diversifying layer API calls. 
To conduct an ablation study of the proposed MCMC-based search algorithm, we constructed a baseline named \memor.
\memor is the same as \toolname, except that \memor adopts a random mutation strategy instead of the MCMC-based search algorithm.
In other words, \memor randomly selects a mutation operator in each iteration for model generation. If \toolname can outperform \memor, our MCMC-based search algorithm would help guide model generation towards higher diversity and test coverage.

\begin{figure}[t!]
    \centering
    \includegraphics[width=1.0\linewidth]{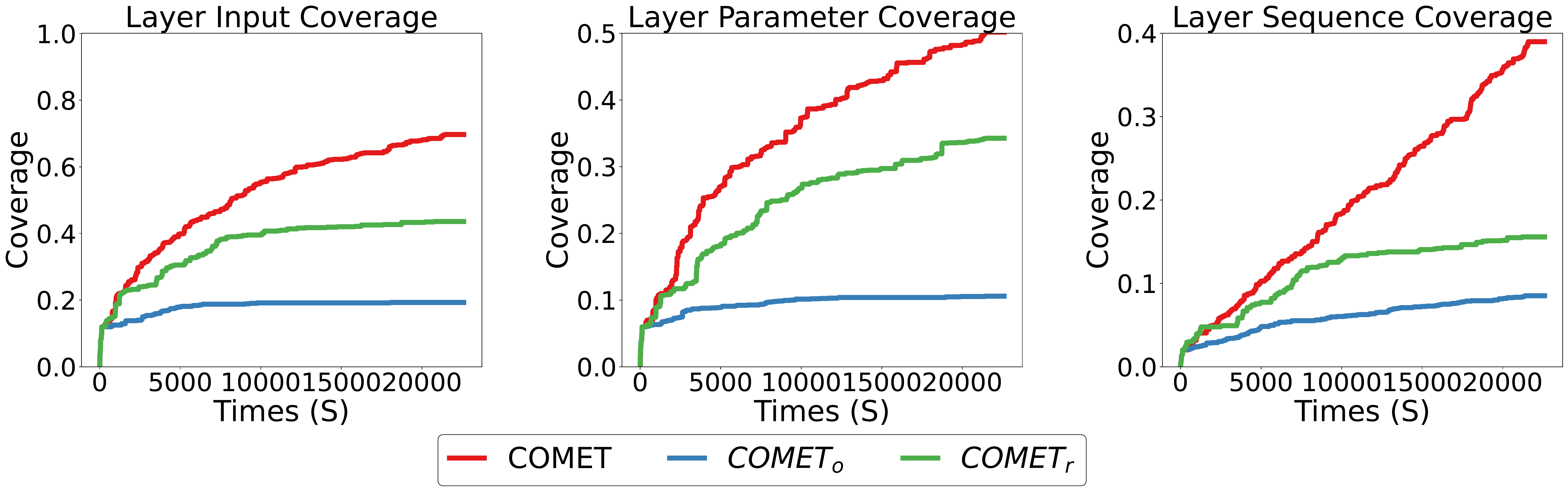}
    \caption{Comparison of \toolname, \memoo, and \memor}
    \label{fig:rq4_ablation}
\end{figure}

We ran \toolname, \memoo, and \memor for six hours. Similar to the setting in \S\ref{eval_simplification}, we use layer input coverage, layer parameter coverage, and layer sequence coverage to evaluate their effectiveness.
We present our results in Figure~\ref{fig:rq4_ablation}. 
Our result shows that, compared to \memoo and \memor, \toolname can effectively increase the layer API call diversity of the generated models on TensorFlow, and the improvement is significant.
In particular, compared with \memoo, \toolname can cover 50.4\% more layer inputs, increase the layer parameter coverage by 39.6\%, and improve the layer sequence coverage by 30.5\%. The improvement by using our proposed search algorithm is also noticeable. Besides improving layer API call diversity, we observe that \toolname's MCMC-based search algorithm also contributes to the test coverage. Specifically, \toolname can increase the branch coverage of \memoo and \memor by 6.40\% and 1.09\%, respectively.

The result of the ablation study shows that both our mutation operators and search algorithm contribute to generating diverse DL models.

%% file: Table-Origin-Model-Statistics.tex
\begin{table}[t!]
\caption{Statistics of Original Models}
\small
\label{tab:origin_model_statistics}
\renewcommand{\arraystretch}{1}
\setlength{\tabcolsep}{2.2pt}
\resizebox{0.9\linewidth}{!}{
\begin{tabular}{l|l|l|l|l}
\hline
\multirow{1}{*}{Model Name} & Total \# Layers (Weights)  & Layer Inputs & Layer Parameters  & Layer Sequences\\ \hline\hline
AlexNet
& 17 (21.592M)
& 22
& 35
& 5\\ \hline
\rowcolor{Gray}
LSTM (SineWave)
& 5 (0.071M)
& 9
& 22
& 2\\ \hline
LeNet
& 9 (0.062)
& 14
& 20
& 3\\ \hline
\rowcolor{Gray}
DenseNet121
& 429 (8.162M)
& 29
& 45
& 6\\ \hline
InceptionResNetV2
& 782 (55.973M)
& 26
& 46
& 4\\ \hline
\rowcolor{Gray}
InceptionV3
& 313 (23.951)
& 30
& 45
& 4\\ \hline
ResNet50
& 177 (25.736M)
& 24
& 40
& 4\\ \hline
\rowcolor{Gray}
Xception
& 134 (23.010M)
& 30
& 54
& 3\\ \hline
LSTM (StockPrice)
& 3 (0.026M)
& 6
& 21
& 1\\ \hline
\rowcolor{Gray}
MobileNetV2
& 156 (3.639M)
& 37
& 54
& 9\\ \hline
\end{tabular}
}
\end{table}

%% file: Table-Evaluation-Baseline.tex
\begin{table}[t!]
\caption{Coverage by \textit{\toolname} and Baselines}
\small
\label{tab:evaluation_baseline}
\renewcommand{\arraystretch}{1.5}
\setlength{\tabcolsep}{2.2pt}
\begin{tabular}{l|l|l|l|l|l}
\hline
\multirow{2}{*}{Technique} & Branch  & Line & Layer Input & Layer Parameter  & Layer Sequence\\ 
& Coverage & Coverage & Coverage & Coverage & Coverage\\ \hline\hline
\toolname
& \textbf{19.4\%}
& \textbf{34.9\%}
& \textbf{69.7\%}
& \textbf{50.2\%}
& \textbf{39.0\%}\\ \hline
\rowcolor{Gray}
MUFFIN
& 16.0\%
& 30.4\%
& 34.1\%
& 25.9\%
& 15.6\%\\ \hline 
LEMON
& 12.5\%
& 26.6\%
& 14.2\%
& 6.4\%
& 2.4\%\\ \hline
\rowcolor{Gray}
CRADLE
& 12.0\%
& 25.9\%
& 10.4\%
& 5.9\%
& 1.2\%\\ \hline
\end{tabular}
\end{table}

%% file: Table-Evaluation-Bug.tex
\begin{table*}[t!]
\renewcommand{\arraystretch}{1.5}
\caption{Performance on Bug Detection}
\resizebox{1.0\linewidth}{!}{
\begin{tabular}{l|l|l|l|l|l|l|l|l}
\hline
& \multicolumn{8}{c}{DL Libraries} \\ \hline
& ONNXRuntime   & MXNet        & Keras-MXNet    & TF2ONNX       & ONNX2PyTorch  & Keras & TensorFlow & PyTorch\\ \hline \hline

Crash &
2 (2) &
4 (0) &
1 (0) &
4 (3) &
4 (1) &
8 (7) &
0 (0) &
0 (0) \\\hline
NaN &
2 (2) &
1 (1) &
0 (0) &
1 (1) &
0 (0) &
4 (3) &
0 (0) &
0 (0) \\\hline
Inconsistency &
0 (0) &
0 (0) &
0 (0) &
0 (0) &
0 (0) &
0 (0) &
1 (1) &
0 (0) \\\hline \hline
Subtotal  &
4 (4) &
5 (1) &
1 (0) &
5 (4) &
4 (1) &
12 (10) &
1 (1) &
0 (0) \\\hline \hline
\end{tabular}
}
\begin{tablenotes}
  \small
  \item The number of confirmed bugs is parenthesized.
\end{tablenotes}
\label{tab:evaluation-bug}
\end{table*}

%% file: Table-Discussion-Model-Chracteristics.tex
\begin{table}[htbp!]
\caption{Analysis On Fault-Triggering Conditions of Bugs Detected By \toolname 
}
\small
\label{tab:discussion_memo_bugs}
\renewcommand{\arraystretch}{1.5}
\setlength{\tabcolsep}{2.2pt}
\resizebox{0.8\linewidth}{!}{
\begin{tabular}{l|l|l|l|l|l|l}
\hline
\multirow{1}{*}{Library} & Issue ID & Layer Input & Layer Parameter & Layer Sequence  & Single Layer W/ Default Parameter & Other \\ \hline\hline
ONNXRuntime
& 11173
&
& 
& \cmark
& 
& \cmark\\ \hline
\rowcolor{Gray}
ONNXRuntime
& 11024
& \cmark
& 
& 
& 
& \\ \hline
ONNXRuntime
& 11010
&  
&  
&  
&  
& \cmark\\ \hline
\rowcolor{Gray}
ONNXRuntime
& 11241
&  
&  
& \cmark
&  
&  \\ \hline
MXNet
& 20947
&  
& \cmark
&  
&  
& \cmark\\ \hline
\rowcolor{Gray}
Keras
& 16348
&  
& \cmark
&  
&  
&  \\ \hline
Keras
& 16273
&  
& \cmark
& \cmark
&  
&  \\ \hline
\rowcolor{Gray}
Keras
& 16158
&  
&  
&  
&  
& \cmark\\ \hline
Keras
& 16289
&  
&  
&  
&  
& \cmark\\ \hline
\rowcolor{Gray}
Keras
& 16314
&  
& \cmark
&  
&  
&  \\ \hline
TF2ONNX
& 1875
&  
& \cmark
&  
&  
& \cmark\\ \hline
\rowcolor{Gray}
TF2ONNX
& 1807
&  
&  
&  
& \cmark
&  \\ \hline
ONNX2PyTorch
& 33/34
&  
&  
& \cmark
&  
&  \\ \hline
\rowcolor{Gray}
TF2ONNX
& 1811
&  
&  
& \cmark
&  
&  \\ \hline
Keras
& 16314
&  
& \cmark
&  
&  
&  \\ \hline
\rowcolor{Gray}
Keras
& 16492
& \cmark
&  
&  
&  
& \cmark\\ \hline
Keras
& 16927
& \cmark
&  
&  
&  
&  \\ \hline
\rowcolor{Gray}
Keras
& 16933
& \cmark
& 
&  
&  
&  \\ \hline
Keras
& 16970
& \cmark
& \cmark 
&  
&  
&  \\ \hline
\rowcolor{Gray}
Keras
& 17044
& \cmark
& 
&  
&  
&  \\ \hline
TF2ONNX
& 1903
& 
&  
& \cmark 
&  
&  \\ \hline
Count
& 
& 6
& 7
& 6
& 1
& 7\\ \hline
\end{tabular}
}
\end{table}

%% file: Table-Evaluation-Reduction.tex
\begin{table}[t!]
\caption{Performance of Model Synthesis Algorithm}
\small
\label{tab:evaluation_synthesis}
\renewcommand{\arraystretch}{1.5}
\setlength{\tabcolsep}{2.2pt}
\resizebox{1.0\linewidth}{!}{
\begin{tabular}{l|l|l|l|l}
\hline
\multirow{1}{*}{Initial Seed Models} & Total \# Layers (Weights)  & Layer Inputs & Layer Parameters  & Layer Sequences\\ \hline\hline
Original Models
& 2025 (305.28M)
& 116
& 122
& 25\\ \hline
\rowcolor{Gray}
Synthesized Models
& 317.8$\pm$3.9( 104.16K/$\pm$46.44K)
& 123.75$\pm$1.89
& 153.85$\pm$1.53
& 45.25$\pm$2.23\\ \hline
\end{tabular}
}
\end{table}

%% file: 7-Discussions.tex
\section{Discussions}

\subsection{Analysis Of Uncovered Branches}
\label{sec:discuss_uncovered}
As is shown in \S\ref{subsec:rq1}, all existing techniques and \toolname can cover only a portion of branches (at most 19.4\%) and lines (at most 34.9\%) inside TensorFlow. Following the Reachability, Infection, Propagation (RIP) model~\cite{ammann2016introduction}, a program failure requires the location of the program that contains the fault must be reached. In this section, we collected and analyzed some library branches that \toolname cannot cover. Through this study, we hope to better understand why these branches are not reached and how we can reach them in the future. In particular, we collected uncovered branches of two representative modules: {\mycode keras/convolution.py} and {\mycode ops/array\_ops.py}. The former module implements the Keras' layer APIs, and the latter supports implementing low-level tensor operators. Specifically, \toolname covers 173 out of 356 branches in {\mycode keras/convolution.py} and 133 out of 462 branches in {\mycode ops/array\_ops.py}.

We observed two main reasons \toolname fails to cover branches in our analyzed modules. The first reason is caused by the implementation. In line with existing works, we use Keras to generate DL models. However, Keras only provides APIs to construct high-level layers, while using Keras' APIs cannot directly access many low-level operators. As a result, these low-level operators' input diversity is limited. For instance, we found that a specific parameter (i.e., {\mycode axis}) of the low-level operator {\mycode tf.stack} is always {\mycode 0} when using Keras to generate models, resulting in around half of the branches inside this operator's implementation being uncovered. Nevertheless, \toolname can still improve the branch coverage of the state-of-the-art technique (i.e., Muffin) from 16.0\% to 19.4\%, regardless of this implementation restriction. The testing paradigm causes another reason. We observe that in our analyzed modules, there exist some utility functions that cannot be invoked through a DL model. For instance, some private methods in {\mycode Convolution} class are designed for developers, and these methods cannot be invoked through public APIs provided by TensorFlow developers. 

\subsection{Effect Of the Hyperparameter $\sigma$ on Numeric Parameter Mutation}
\begin{figure}[htbp!]
    \centering
    \includegraphics[width=1.0\linewidth]{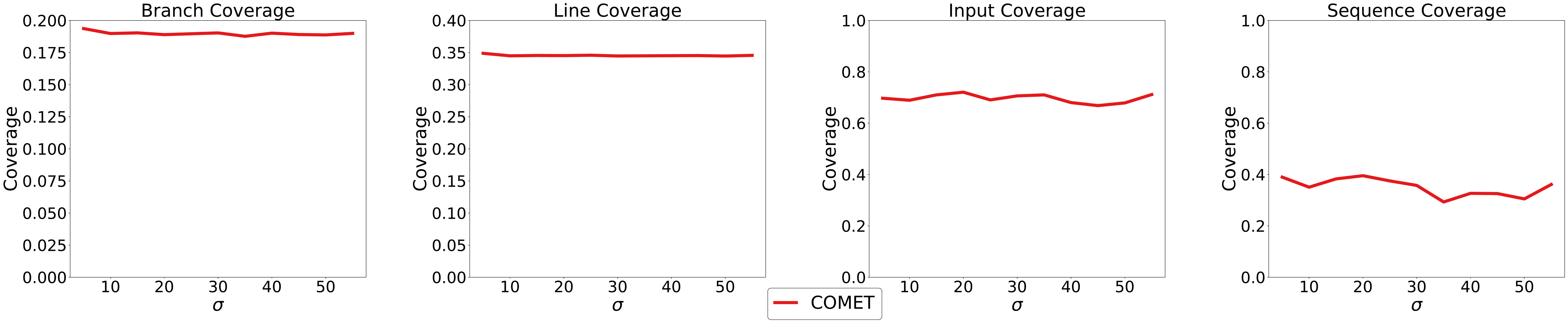}
    \caption{Effect Of Hyperparameter $\sigma$}
    \label{fig:sigma_discussion}
\end{figure}

On top of the remarkable results achieved by \toolname, we explore a further question: \textit{whether the setting of the hyperparameter $\sigma$ defined in \S3.2.2 will affect the effectiveness of \toolname}? 
Specifically, $\sigma$ is used to define the size of value space for numeric parameters so \toolname will not exhaust all possible values. 
In our experiment, we set $\sigma=5$ based on our experience. 
We further discuss whether changing the value of $\sigma$ will influence the model diversity and test coverage on the DL library achieved by \toolname. 
To answer this question, we set $\sigma$ from 5 to 55 with the step of 5, and examined the branch coverage, line coverage, layer input coverage, and layer sequence coverage. We also fix the value range of each numeric parameter for different values of $\sigma$.~\footnote{See our implementation at \url{https://github.com/maybeLee/COMET} for details.} Since changing the value of $\sigma$ will influence the size of test space for numeric parameters, different $\sigma$ will output different layer parameter coverage results on the same DL model. Therefore, we did not compare the layer parameter coverage in this discussion.
We can see from Figure~\ref{fig:sigma_discussion} that branch, line, and input coverage are relatively insensitive to the value of $\sigma$, while sequence coverage may vary a bit but not significantly when we chose different values of $\sigma$. For instance, the branch coverage achieved by \toolname with different $\sigma$ fluctuates from 18.8\% to 19.4\%, the layer sequence coverage fluctuates from 29.3\% to 39.5\%.

%% file: 8-Threats.tex
\subsection{Threats to Validity}
\label{sec:threats}
Overall, there are three threats that may affect the experimental results.
We ran experiments with a time budget of six hours for each baseline on the same machine in turn. Yet, the amount of available system resources for the experiments may vary across 24 hours. 
To alleviate the influence caused by system computing performance, we set the total number of threads/GPU cards used in each experiment to be the same. The second threat is the number of new bugs detected by \toolname. For those unconfirmed crash bugs, following existing work~\cite{lemon}, we used the stack trace to distinguish them. However, it is possible that one crash bug can have different stack traces. To alleviate this problem, we manually checked the root cause of each unconfirmed crash bug to reduce false positives. 
The third threat is introduced by our test coverage measurement. The fact that coverage is measured in only one DL library may affect the evaluation result. To alleviate the bias of measurement, our criteria for selecting the representative library to collect code coverage follow the practice of one recent JVM testing work~\cite{classfuzz}. Specifically, we choose TensorFlow as the representative because 1) it is the DL library with the highest GitHub stars and is the most popular one with the richest implementations, 2) all existing DL library testing techniques consider TensorFlow as their subject under test, and 3) the testing practices and bugs inside DL libraries share a significant commonality~\cite{jia2021unittestquality, oracleapproximation, empirical_2022_dlframeworks, chen2022toward}. Therefore, the code coverage in TensorFlow can reflect DL library testing techniques' effectiveness in terms of test coverage. To collect the branch coverage and line coverage on TensorFlow's model construction and execution modules, we iterated all modules inside TensorFlow's source code and chose the related ones. Indeed, this may introduce some bias (e.g., some modules may be missed by us). To alleviate this problem, we carefully read TensorFlow's documentation and the description in each module. We publish our collected module names on our project site.

%% file: 9-RelatedWork.tex
\section{Related Works}

\subsection{Testing of DL Library}

Besides testing DL libraries by DL models~\cite{cradle,audee,lemon,graphfuzz,muffin}, there are other studies on a related testing problem with different focuses.
DocTer~\cite{docter} extracts the argument constraint of DL library functions from the documentation of DL libraries. To analyze the free-form API documentation, DocTer first preprocesses and parses the documentation into the dependency parse tree and employs a learning algorithm to extract constraints from this documentation. Based on these constraints, it randomly generates two types of test inputs: valid inputs and invalid inputs.
FreeFuzz~\cite{freelunch} also targets collecting DL library functions constraint. Instead of statically collecting the constraints as DocTer, it instruments the library functions and obtains their input constraints by running code/model from open source (i.e., library documentation, unit test suite written by developers, and DL models in the wild). After collecting input constraints, it randomly generates test inputs following them for DL library testing.
Predoo~\cite{predoo} proposes a fuzzing technique to test the operators in the DL library. It aims to estimate and detect individual DL operator's precision errors. Specifically, Predoo collects publicly available test inputs as the seeds and generates variants by iteratively adding perturbations such as Gaussian noise to the initial seeds to maximize output precision errors.
Furthermore, TVMFuzz~\cite{tvmfuzz} focuses on the testing of the DL compilers. 
The major difference between our works and these works is that \toolname generates DL models as test inputs to test DL library modules related to a DL model's execution.

\subsection{Empirical Studies for DL Libraries}
To date, researchers also conducted some empirical studies on DL libraries. We briefly group them into two categories: studies on DL framework bugs and studies on DL library testing practice.
Jia et al.~\cite{bugsinsidetensorflow} report an empirical study to understand the characteristics of the bugs in TensorFlow. They further study the symptoms, causes and repair patterns of TensorFlow bugs~\cite{JIA2021110935}. Their studies reveal that the root causes of a large portion of reported bugs reside in the algorithm implementations or the interfaces (i.e., API) provided by TensorFlow.
Tambon et al.~\cite{silentbugs} study silent bugs such as performance or accuracy issues inside TensorFlow. Silent bugs will lead to incorrect behavior, but they do not cause system crashes or hang, nor show an error message to users. This study indicates that inaccurate calculation result is the most common symptom of silent bugs.
Chen et al.~\cite{chen2022toward} conduct a comprehensive study to facilitate a sufficient understanding of DL library bugs. Compared with previous studies, they took not only TensorFlow but also PyTorch, MXNet, and Deeplearning4j as the study subjects. Based on their analysis, they point out that components implemented for deep learning algorithms are the buggiest ones, and over 53.75\% of bugs are observed at the model training stage, which could lead to lengthy testing and debugging.

Several studies have been proposed to study the DL library testing practice.
Nejadgholi et al.~\cite{oracleapproximation} conduct the first study on the design and evolution of oracle approximation(OA) assertions inside the DL libraries' test suite. They point out that a non-negligible portion (25\%) of OA assertions in DL libraries and developers frequently change OA assertions.
Wang et al.~\cite{wang2021automatic} conduct an empirical study on five machine learning libraries with two popular unit test case generation tools, i.e., EVOSUITE~\cite{evosuite} and Randoop~\cite{randoop}. They find out that most machine learning libraries do not maintain a high-quality test suite to achieve satisfactory code coverage; although existing unit test case generation tools can help improve test suite quality, the improvement is limited.
Jia et al.~\cite{jia2021unittestquality} evaluate the quality of unit test suites in DL frameworks using mutation analysis. They find out that existing test cases are more effective in detecting bugs inside incorrect control flow logic, while they perform worse when wrong values cause the bug.
Instead of focusing on the root causes and symptoms of library bugs or analyzing the testing practice adopted by DL library developers, we focus on the generation of diverse DL models in terms of their layer API calls. 

\subsection{Mutation Testing for Deep Learning}
There are also some mutation testing works proposed to modify DL models. DeepMutation~\cite{deepmutation} proposes a set of source-level mutations to inject faults into the training data and training program, and model-level mutations to inject the faults into the trained DL model by mutating the value of neurons or adding/removing some layers. By doing so, DeepMutation can inject faults into the DL application and use the mutated DL application to evaluate the quality of DL application's test suite (i.e., testing data). DeepCrime~\cite{deepcrime} defines 35 source-level mutation operators based on real DL program faults to inject coding errors into the training programs and labeling errors into the training data. Same as DeepMutation, DeepCrime is also designed to inject faults into the DL application to evaluate the quality of test suites. Wang et al.~\cite{adversamplemutation} apply model-level mutation to inject the faults into the trained DL model by mutating the value of neurons. They further guide adversarial test input generation using the mutated DL model based on the insight that adversarial test input would be more sensitive to mutations on the DL model. Different from these mutation testing works, our work designed mutation operators to generate test input instead of generating the mutants of the subject under test. In other words, we use the mutation operator to apply changes to the seed DL model rather than injecting faults to the subject under test.

%% file: 10-Conclusion.tex
\section{Conclusion}
In this paper, we propose a novel technique named \toolname to test DL libraries by generating diverse layer API calls as test inputs. Motivated by the model generation script of Keras, \toolname defines three coverage criteria to measure the layer API call diversity: layer input coverage, layer parameter coverage, and layer sequence coverage. Driven by these criteria, \toolname proposes a set of novel mutation operators and a coverage-guided search algorithm to search for DL models with diverse layer API calls to achieve more comprehensive testing on DL libraries. It also proposes a model synthesis method to significantly boost the efficiency of diverse model generation by addressing the challenge of runtime overhead incurred by large models. Our evaluation of popular DL libraries shows that \toolname significantly outperforms the state-of-the-art DL library testing techniques in the effectiveness of diverse model generation. In total, \detectedbugs new bugs were detected by \toolname, among which, \confirmedbydevelops of them were confirmed by developers and \fixedbugs of these confirmed bugs have been fixed by developers.

%% file: 11-Acknowledgements.tex
\section*{Acknowledgements}
We would like to thank the anonymous reviewers for their comments and suggestions. We would also like to thank DL library developers for analyzing our reported issues. This work was supported by the National Natural Science Foundation of China (Grant Nos. 61932021, 62002125), the National Key Research and Development Program of China (Grant No. 2019YFE0198100), the Hong Kong RGC/GRF (Grant No. 16205722), the Hong Kong ITF (Grant No. MHP/055/19), and the MSRA Collaborative Research Grant.